\documentclass[preprint,1p,authoryear]{elsarticle}

\usepackage{amssymb}
\usepackage{todonotes}

\usepackage{amsmath}
\usepackage{subfig}
\usepackage{mwe}
\usepackage{multirow}
\usepackage{url}
\usepackage{fancyhdr}
\usepackage{tabu}
\usepackage{booktabs}
\usepackage{xparse}

\NewDocumentCommand{\rot}{O{45} O{1em} m}{\makebox[#2][l]{\rotatebox{#1}{#3}}}%

\journal{Journal of Parallel and Distributed Computing
}

\begin{document}

\begin{frontmatter}



\title{Scalable System Scheduling for HPC and Big Data}


\author{Albert Reuther*, Chansup Byun, William Arcand, David Bestor, Bill Bergeron, Matthew Hubbell, Michael Jones, Peter Michaleas, Andrew Prout, Antonio Rosa, Jeremy Kepner}
\cortext[cor1]{Corresponding Author, Email:  reuther@mit.edu; Address: MIT Lincoln Laboratory Supercomputing Center, 244 Wood Street, Lexington, MA 02421, USA  }
\cortext[cor2]{This material is based upon work supported by the National Science Foundation under Grant No. DMS-1312831.  Any opinions, findings, and conclusions or recommendations expressed in this material are those of the authors and do not necessarily reflect the views of the National Science Foundation.}

\address{Massachusetts Institute of Technology}

\begin{abstract}
In the rapidly expanding field of parallel processing, job schedulers are the ``operating systems'' of modern big data architectures and supercomputing systems.  Job schedulers allocate computing resources and control the execution of processes on those resources.  Historically, job schedulers were the domain of supercomputers, and job schedulers were designed to run massive, long-running computations over days and weeks.  More recently, big data workloads have created a need for a new class of computations consisting of many short computations taking seconds or minutes that process enormous quantities of data.  For both supercomputers and big data systems, the efficiency of the job scheduler represents a fundamental limit on the efficiency of the system.  Detailed measurement and modeling of the performance of schedulers are critical for maximizing the performance of a large-scale computing system.  This paper presents a detailed feature analysis of 15 supercomputing and big data schedulers.  For big data workloads, the scheduler latency is the most important performance characteristic of the scheduler.  A theoretical model of the latency of these schedulers is developed and used to design experiments targeted at measuring scheduler latency. Detailed benchmarking  of four of the most popular schedulers (Slurm, Son of Grid Engine, Mesos, and Hadoop YARN) are conducted.  The theoretical model is compared with data and demonstrates that scheduler performance can be characterized by two key parameters: the marginal latency of the scheduler $t_s$ and a nonlinear exponent $\alpha_s$.  For all four schedulers, the utilization of the computing system decreases to \textless 10\% for computations lasting only a few seconds.  Multi-level schedulers (such as LLMapReduce) that transparently aggregate short computations can improve utilization for these short computations to \textgreater 90\% for all four of the schedulers that were tested.
\end{abstract}


\end{frontmatter}


\section{Introduction}
\label{sec:Intro}

The evolution of computing into the hyperscale data center era has created massive computing infrastructures that can be harnessed to solve a wide range of problems.  The efficient utilization of these resources is critical to their economical and environmental viability.  A 100-megawatt data center that wastes even 1\% of its computing cycles can nullify all the energy-saving measures of a small city.   The operating system of a data center is the job scheduler that decides which tasks are executed on which resources.
Multiple users submit a wide variety of computational jobs to be processed on computational resources that include various (and sometimes heterogeneous) processing cores, network links, memory stores, storage pools, and software execution licenses (as necessary to run licensed applications). Further, each user's job will follow different parallel execution paradigms from independent process jobs to independently (pleasantly) parallel to synchronously parallel jobs, each of which imposes certain execution requirements. In other words, heterogeneity is realized in multiple dimensions. Job schedulers are a key part of modern computing infrastructures; if the scheduler is not effectively managing resources and jobs, then the computing capabilities will not perform well.

Job schedulers go by a variety of names, including schedulers, resource managers, resource management systems (RMS), and distributed resource management systems (D-RMS). These terms are used interchangeably. The two main terms -- scheduler and resource manager -- capture the primary activities of this type of distributed software. At their simplest level, job schedulers match and execute compute jobs from multiple users on multiple computational resources. The users and their jobs usually have different resource requirements and priorities. Similarly, the computational resources have different capabilities and availabilities, and these change dynamically as different jobs are executed. Further, the resources must be managed in such a way that they are best utilized, given the mix of jobs that need to be executed.

\begin{figure}[htb]
    \centering
    \includegraphics[width=20pc]{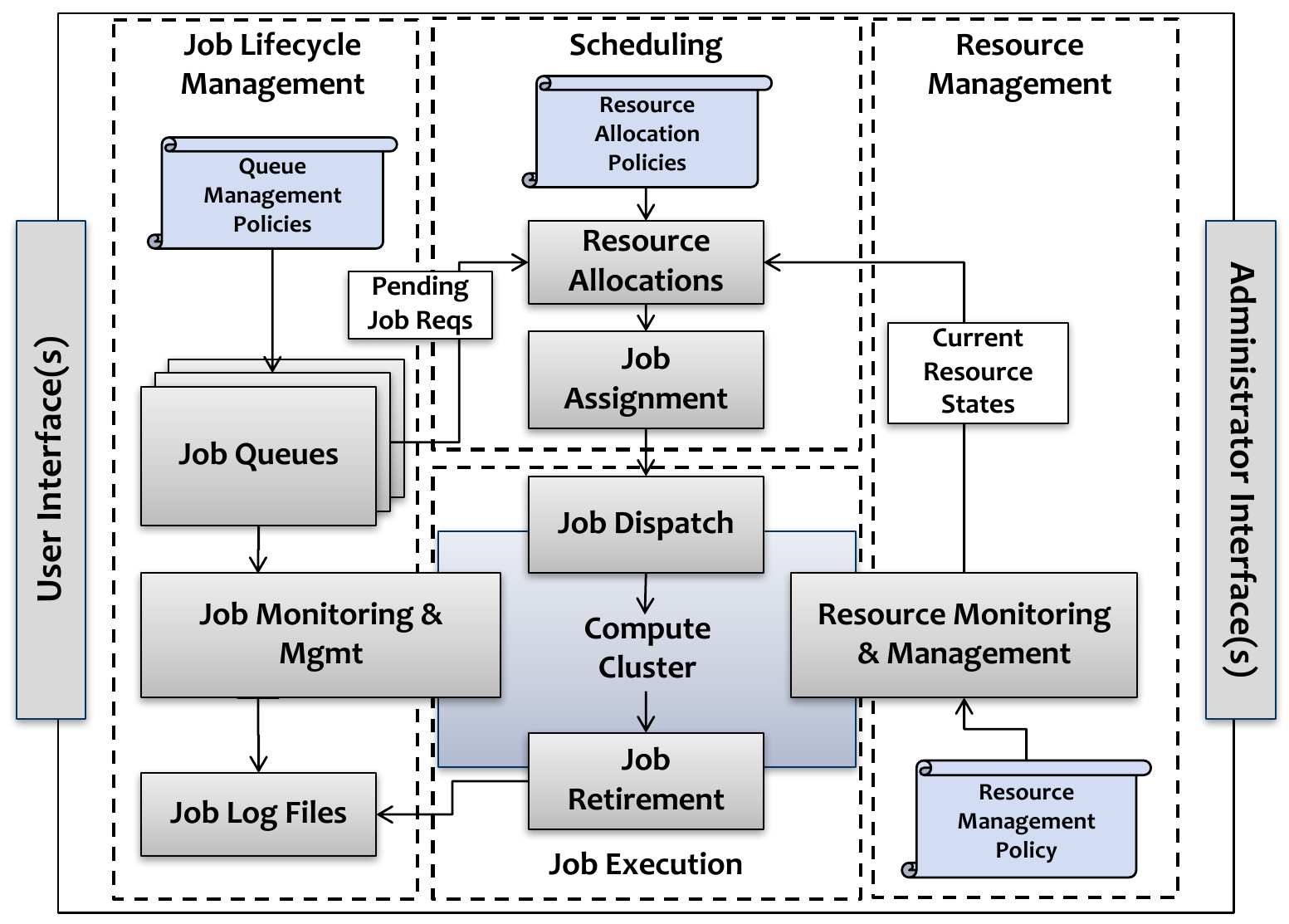}
    \caption{Key components of cluster schedulers, including job lifecycle management, resource management, scheduling, and job execution.}
    \label{fig:SchedulerArchitecture}
  \end{figure}

In the functional model of schedulers, every job scheduler has an architecture with four key functions: job lifecycle management, resource management, scheduling, and job execution.  These functions are shown in Figure~\ref{fig:SchedulerArchitecture} and are described in greater detail as follows:

\begin{itemize}
\item The job lifecycle management function receives jobs from users through the user interface and places them in the job queue(s) to wait for execution (even if other jobs are already scheduled). Various resources for the job, including memory, licenses, and accelerators (such as GPUs), are requested through the user interface by the user. The job lifecycle management task is also responsible for prioritizing and sorting candidate jobs for execution by using the queue management policies. 
\item The resource management function receives availability and resource state information from the compute nodes, aggregates it, and makes it available to the scheduler. It also collects job status information to make available to users and to record in logs. 
\item The scheduling function allocates resources (including one or more job slots on compute nodes, compute node local memory, accelerator cards, etc.) and assigns the job to the resource(s) if the resources are available to run the job. 
\item The job execution function is responsible for dispatching/launching the job on the resources. Upon the completion of each job, the job execution function manages closing down the job and sending the job statistics to the job lifecycle management function, which records it in logs. 
\end{itemize}

\begin{figure}[htb]
    \centering
    \includegraphics[width=20pc]{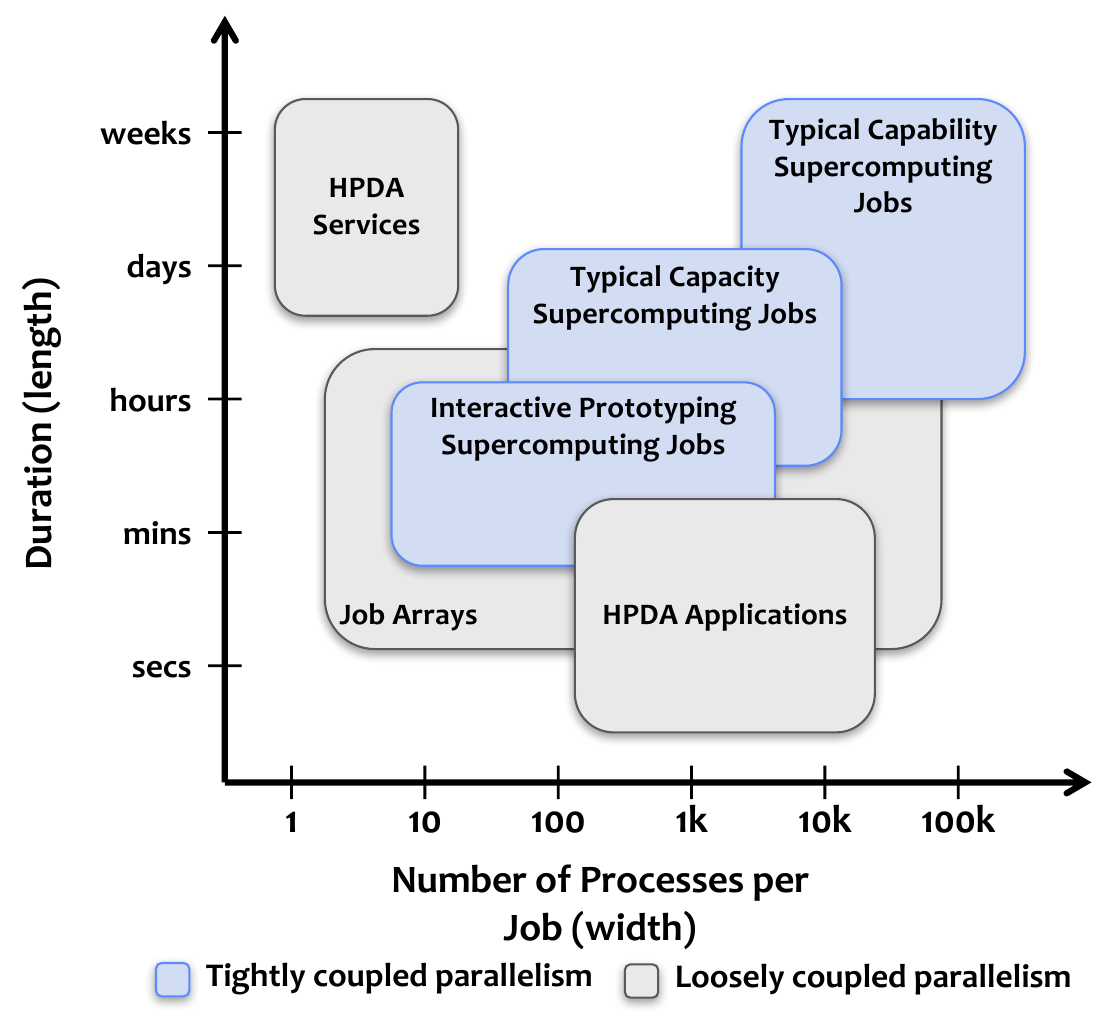}
    \caption{Characterization of HPC and big data jobs in terms of execution time and parallelism.}
    \label{fig:JobTypes}
  \end{figure}

This architecture has stood the test of time; it is not much different from those first developed for parallel computing schedulers [\cite{jones1996requirements}, \cite{saphir1995job}]. However, there has been significant evolution in the sophistication of resource management, scheduling, resource allocation, and queue management as resources and jobs have become more diverse. This greater heterogeneity can be captured in a number of dimensions, the most basic being execution time and parallelism. Figure~\ref{fig:JobTypes} depicts the characterization of job types from both the traditional high performance computing and big data paradigms. Traditional HPC jobs can be categorized into single-process jobs, a set of single-process jobs that are grouped together and executed asynchronously (job arrays), and sets of processes that are all launched and executed simultaneously and synchronously (parallel jobs).
Interactive prototyping parallel jobs have many of the same characteristics of typical capacity supercomputing jobs, but they tend to occupy fewer processors and execute for less time. Most HPC systems are configured to submit jobs to a batch queue, in which they wait to be scheduled from minutes to days. However, there are mechanisms, such as interactive queues, immediate execution flags, and slot reservations, which enable interactive job execution. Big data jobs tend to be in two categories. Some big data jobs are services jobs in which one or a few processes are scheduled and executed to provide services to other jobs. The majority of big data jobs are very short-run, time-critical jobs that execute in a matter of seconds to minutes and are independent of each other. Big data jobs are expected to execute immediately; that is, they tend not to wait in batch queues. High performance data analytics jobs have characteristics of both HPC and big data types of jobs. This paper explores the features that are necessary to execute high performance data analytics jobs because we need the best of both worlds to best accommodate such jobs. 

This paper makes the following contributions: 
\begin{itemize}
\item It introduces the categorization of schedulers into scheduler families and categorizes a set of HPC and big data into these families. 
\item It enumerates a set of scheduler features that are important for high performance data analysis jobs and compares the support for these features for a representative set of HPC and big data schedulers.
\item It recognizes that many features are supported across the schedulers but that a key scheduler feature that has not been explored in much detail is job scheduler latency. 
\item It derives a mathematical description for job launch latency.
\item It presents a benchmarking strategy for measuring job launch latency and benchmarks a representative set of HPC and big data schedulers.
\item It measures the marginal latency and nonlinear exponent of job launch latency of the benchmark results of the representative set of HPC and big data schedulers. 
\end{itemize}

The outline of the article is as follows.  Section~\ref{sec:RelatedWork} surveys related work. Section~\ref{sec:Schedulers} overviews the most prominently used schedulers, organizes them into scheduler families (schedulers with similar features), and compares their features. Section~\ref{sec:Modeling} introduces latency and utilization models that are important for understanding the job launch latency of schedulers.  Section~\ref{sec:Performance} introduces a job- and resource-agnostic job scheduling and execution latency performance benchmark and presents the performance results of several representative schedulers along with a discussion of the results. Section~\ref{sec:summary} summarizes the paper.


\section{Related Work}
\label{sec:RelatedWork}


In the world of schedulers, a great deal of research has gone into various components, algorithms, and implementations of job schedulers. Much research has been published on optimizing the placement of jobs onto the resources with a great deal of variety in both the homogeneity and heterogeneity of the jobs and the resources. There has also been significant research in how to characterize and model the execution and resource utilization of various algorithms and applications, both as serial jobs and parallel jobs. A vast number of papers, book chapters, and books have been published on these topics; this Related Work section will overview the most significant literature to differentiate our work on job scheduler features and performance to enable high performance data analytics jobs. 

First, it is worthwhile to address the related work in survey and comparison literature. This literature addresses all four functional areas of schedulers, while the feature comparisons most closely address the job lifecycle management functionality. There have been quite a number of such papers, and they were often published as final reports of studies conducted to choose a new scheduler for an organization's supercomputing and/or big data systems. The first papers that brought about comparisons of schedulers came from NASA in the late 1990s when they were replacing the Network Queuing System (NQS) batch scheduler at NASA [\cite{kingsbury1986network}]. This work prompted two reports that catalyzed the development of the Portable Batch System (PBS)~[\cite{jones1996evaluation}, \cite{jones1997second}]. Feitelson and Rudolph's 1995 paper [\cite{feitelson1995parallel}] established a basis for making comparisons between job schedulers, and as other schedulers were developed by other organizations, more such comparison papers were written,  including~\cite{baker1996review}, \cite{byun2000comparison}, \cite{el2004performance}, and \cite{yan2008comparative}. In the early 2000s, a number of universities and research organizations were developing technology to share supercomputing resources across their organizations as computing grids~[\cite{foster2003grid}]. Among the significant papers and books from the research of scheduling for grid computing were \cite{czajkowski1998resource}, \cite{krauter2002taxonomy}, and \cite{nabrzyski2004grid}. 
Similarly among big data schedulers, there have been comparisons between Mesos, MapReduce, and Hadoop YARN~[\cite{xavier2014performance}] and between several Google schedulers~[\cite{burns2016borg}]. 
More recently, there have been a few articles that make comparisons between traditional HPC schedulers, grid schedulers, and big data schedulers~[\cite{getov2015workload}], as well as scheduling in cloud computing~[\cite{hussain2013survey}, \cite{singh2016survey}]. 
In all of these comparison papers, the comparison examines some subset of the features that were considered in service to certain processing workloads, usually real workloads or stochastic models of real workloads. Further, the comparisons have been within the traditional HPC or big data job scheduling areas, but have not taken candidate schedulers from both. 


Next we must address related work within the scheduling function. There are a number of papers that focus on comparing the queue management and scheduling algorithms within schedulers, including discussion of support for first-come, first-served (FCFS)~[\cite{schwiegelshohn1998analysis}], fairshare~[\cite{schwiegelshohn2000fairness}], backfill~[\cite{lifka1995anl}, and gang scheduling~[\cite{ousterhout1982scheduling}, \cite{feitelson1997improved}]. These papers includes comparisons of how jobs are prioritized within queues and how jobs are matched to resources to most effectively execute them. 
These articles make comparisons among the various schedulers available within Apache Hadoop YARN~[\cite{lin2016performance}]; within PBS~[\cite{henderson1995job}], within the Maui scheduler~[\cite{jackson2001core}, and among several of the HPC schedulers~[\cite{feitelson2004parallel}, \cite{etsion2005short}, \cite{yan2008comparative}]. 

The publications that are covered in the previous paragraphs are all comparisons of actual schedulers as well as the application of queue management and scheduling algorithms on actual workload traces and simulated workload traces based on actual workload characteristics. However, another branch of job scheduler research has approached this topic from a mathematical and statistical operations research vector. 
Statistical distributions are applied to various aspects of workload characteristics and resource characteristics. First, a taxonomy of workloads, resources/platforms, and mapping and scheduling heuristics is presented in~\cite{ali2005characterizing}. Workload modeling is extensively treated in~\cite{feitelson2002workload}, while~\cite{leung2004handbook} covers a very large set of scheduling algorithms. An extensive study and comparison of mapping and scheduling algorithms was reported in~\cite{braun2000comparison}, with refinements in~\cite{braun2008static}. Further refinements of the underlying statistical models and parameter set generation was published in~\cite{canon2015heterogeneity} and~\cite{canon2015heterogeneity_tech_report}. 


Many of these studies have put much work into crafting a representative mix of heterogeneous resources (i.e., memory, GPUs, cores, compute power) that best represent the systems they are modeling and a representative mix of job workloads that their systems execute. However, these studies primarily stress the ability of the schedulers to match jobs to resources and only test the scheduling and matching efficiency of the schedulers. While that research is extremely important to the effectiveness of job schedulers, job execution is often most important to the short serial and parallel data analytics jobs.

There are several papers that have specifically addressed the efficiency of launching jobs, i.e., measureing the scheduling and job execution functions in schedulers.
Two papers compare job launch time of a monolithic (single) scheduler to that of distributed/partitioned job scheduling. \cite{brelsford2013partitioned} explores partitioned parallel job scheduling by modifying the IBM LoadLeveler (a modification of HTCondor) scheduler, while \cite{zhou2013exploring} explores distributed resource-allocation techniques by modifying Slurm. Rather than measuring utilization, both papers measure the throughput of how many jobs they can launch through the scheduler per second. As one might expect, their partitioned parallel and distributed scheduling techniques yielded greater job throughput. 
A white paper by the Edison Group [\cite{elkourie2014hpc}] also uses job throughput to compare several modern HPC schedulers. 
The \cite{el2004performance} paper measures throughput in jobs per hour and turn-around time of four HPC schedulers -- PBS, LSF, Sun Grid Engine/CODINE, and Condor -- for small, medium, and large jobs with low, medium, and high submission rates. El-Ghazawi and colleagues randomized the CPU time of their jobs, and the average CPU times per job were 22 seconds, 7 minutes and 22 seconds, and 16 minutes and 51 seconds for the small, medium, and large jobs, respectively. Further, each job submission experiment only consisted of 150 jobs. Finally, our team measured launch times of virtual machine jobs onto clusters using a variety of job slot-packing and oversubscription techniques [\cite{reuther2012hpc}]. However, we did not make a distinction between the latency of the scheduler and the latency of the virtual machine launching because we were reasonably confident that most of the latency was due to virtual machine launching. Clearly, there is a need for more extensive exploration of job launch efficiency, which we address in Sections~\ref{sec:Modeling} and~\ref{sec:Performance}.

\section{Job Schedulers}
\label{sec:Schedulers}

As mentioned in Section~\ref{sec:RelatedWork}, a comparison among both traditional HPC or big data job schedulers has yet to be conducted. In this section, we do so. First, we organize these traditional HPC and big data job schedulers into several scheduler families to accentuate common intent and development among them. These families also aid in choosing a representative set of eight representative schedulers for comparison. In this comparison of representative schedulers, we find that there are many commonalities across all of the schedulers. However, we also find that there are several key differences, mainly between the traditional HPC schedulers and big data schedulers. 
 
\subsection{Scheduler Families}
\label{subsec:SchedFamilies}

A number of production-quality schedulers have been developed over the past 35 years to address various supercomputing and parallel data analysis computer architectures, network architectures, and software architectures. One of the first full-featured schedulers was the Network Queuing System (NQS) batch scheduler at NASA [\cite{kingsbury1986network}]. From the time of NQS, HPC systems have used job schedulers to manage jobs and resources. These schedulers used a batch queue that kept a backlog of jobs to be run, giving the scheduling function many choices from which to choose the optimal next job on the basis of the current and/or near-future resource availability. However, this batch queue required that jobs wait in the queue, often from minutes to days, in order to begin execution. While most HPC centers continue to use batch queues for scheduling jobs, the MIT Lincoln Laboratory team demonstrated on-demand job launching was possible and desirable for many classes of users [\cite{reuther2005technology}]. Likewise, the big data community, including Google, Yahoo, Microsoft, and others, found batch-queuing schedulers to be insufficient for their needs. Big data jobs often are short in duration, run independently, and persistently perform data analysis. HPC jobs tend to be long-running, synchronously parallel simulation and modeling jobs. 

The MapReduce scheduler, which was among the first big data schedulers, is a very simple scheduler, and it was developed because Google MapReduce [\cite{dean2008mapreduce}] (and consequently Hadoop MapReduce [\cite{dittrich2012efficient}]) needed a scheduler that could dynamically allocate jobs only to compute nodes that were storing a local copy of the required input files. Subsequently, schedulers like Borg [\cite{verma2015large}] and Mesos [\cite{hindman2011mesos}] were developed because of the impression that HPC schedulers were designed to be used in batch processing modes; that is, their main feature was the capability of optimal scheduling of jobs in batch queues. However, both of these shifts also occurred because of the need for in-language APIs beyond just a command line interface. There was an effort to develop a common in-language API called DRMAA (Distributed Resource Management Application API) for batch schedulers, but the adoption of DRMAA was not widespread because of nontechnical market factors.



Hence, the main division is between HPC and big data schedulers. The HPC scheduler family can be further broken into the traditional HPC and the new HPC sub-families. The traditional HPC schedulers include PBS [\cite{henderson1995job}], Grid Engine [\cite{slapnivcar2001resource}], HTCondor [\cite{litzkow1988condor}], OAR [\cite{capit2005batch}], and LSF [\cite{zhou1993utopia}]. The new HPC schedulers include Cray ALPS [\cite{newhouse2006alps}] and Slurm [\cite{yoo2003slurm}]. The big data schedulers can be further broken into commercial and open-source sub-families. The commercial big data schedulers include Google MapReduce [\cite{dean2008mapreduce}], Google Borg [\cite{verma2015large}], and Google Omega [\cite{schwarzkopf2013omega}], and components of Borg and Omega are available as the open-source Kubernetes project [\cite{burns2016borg}]. The open-source big data schedulers include Apache Hadoop MapReduce [\cite{dittrich2012efficient}], Apache Hadoop YARN [\cite{vavilapalli2013apache}], and Apache Mesos [\cite{hindman2011mesos}]. There are many plug-ins available for Apache Hadoop and Apache Hadoop YARN, and the two plug-ins that enhance Hadoop YARN functionality the most are Apache Llama [\cite{cloudera2016llama}] and Apache TEZ [\cite{saha2015apache}]. Finally, there are many research schedulers that have been developed in academia and industry to explore and evaluate new job scheduler concepts and ideas. Two recent schedulers in this category are Pacora [\cite{bird2014optimizing}, \cite{bird2011pacora}] and Sparrow [\cite{ousterhout2013sparrow}].  More details about each of the aforementioned schedulers follow.  

\subsubsection{Traditional HPC Family}
\label{subsubsec:TraditionalHPCFamily}
The Portable Batch System (PBS) Scheduler Family is directly descended from NQS, the first batch scheduler developed under NASA funding. PBS is a fully featured job scheduler that includes a separate queue manager, resource manager, and scheduler, and the Maui scheduler is often used in place of the native PBS scheduler. PBS continues to be developed by Altair Engineering (PBSPro) and Adaptive Computing (TORQUE/Maui/Moab Cluster Suite), which both offer open-source and commercial versions. Recently, scalability challenges have been addressed, and full job array integration has been added. 

Grid Engine is a full-featured, very flexible scheduler originally developed and released under the name CODINE in 1993 by Genias Software [\cite{genias1994codine}], later acquired and matured by Sun Microsystems and Oracle, and currently offered commercially by Univa. There are also several open-source versions, including Son of Grid Engine and Open Grid Scheduler, though further development of these offerings seems to be waning. 

HT Condor (High Throughput Condor) continues to be developed by Prof. Myron Livny's team at the University of Wisconsin. HT Condor was designed to cycle-scavenge from distributed heterogeneous desktop computers to managed clusters. It implements a single queue with many resource- and job-distinguishing options, including prioritization and fair execution. It works especially well for many smaller, single-process jobs with execution on a diverse set of computers, clusters, and supercomputers. IBM LoadLeveler was based on Condor [\cite{loadleveler1993guide}, \cite{kannan2001workload}, \cite{epema1996worldwide}]. 

OAR [\cite{capit2005batch}] was developed by the Laboratory of Informatics in Grenoble, France. It is a full-featured scheduler with queues and resources managed through database records and PERL language modules.  

IBM Platform Load Sharing Facility (LSF) is a full-featured and high performing scheduler that is very intuitive to configure and use. It was based on research and development of the Utopia job scheduler at the University of Toronto [\cite{zhou1993utopia}]. OpenLAVA [http://www.openlava.org/] is an open-source derivative of LSF that has reasonable feature parity.

\subsubsection{New HPC Family}

Cray Application Level Placement Scheduler (ALPS) [\cite{newhouse2006alps}] was developed in the early 2000s to take advantage of unique Cray architecture features. In 2013-14, it was refactored as a ''native'' Slurm implementation. 

The Simple Linux Utility for Resource Management (Slurm) [\cite{yoo2003slurm}] is an extremely scalable, full-featured scheduler with a modern multithreaded core scheduler and a very high performing plug-in module architecture. The plug-in module architecture makes it highly configurable for a wide variety of workloads, network architectures, queueing policies, scheduling policies, etc. It was inspired by the Quadrics RMS scheduler [\cite{yoo2003slurm}]. To ease transition from other schedulers, translation interfaces are provided. Slurm began development in the early 2000s at Lawrence Livermore National Laboratory to address scalability issues in job schedulers. It is entirely open source, while consulting services and community development leadership are provided by SchedMD. 

\subsubsection{Commercial Big Data Family}

Google MapReduce was introduced by Google through the \cite{dean2008mapreduce} paper, which detailed how they implemented a map-reduce processing paradigm to process data files stored locally on each compute node. The job scheduler was a simple first-in, first-out scheduler with few features. Subsequent improvements added more features as Google required greater capabilities. 

Google Borg is a software container management system [\cite{verma2015large}, \cite{burns2016borg}] that evolved from the diversification of workloads at Google that expanded past the map-reduce paradigm of parallel computation. The default execution environment through Borg is Linux containers with static library linking for application resource and execution isolation. The scheduler in Borg is a multithreaded scheduler that accommodates batch and persistent jobs. Further, it is highly scalable; it can manage 10,000 to 100,000 nodes. 

Google Omega (Kubernetes/Pilots) [\cite{burns2016borg}, \cite{schwarzkopf2013omega}] was based on Borg and was written with improved software architecture and software engineering principles. It is a distributed job scheduler, and it decouples the cluster state from the resource management and scheduling functionalities for greater scalability and flexibility. Some of these features, including distributed scheduling services, were rolled back into Borg. 

The Apollo scheduler~[\cite{boutin2014apollo}] developed by Microsoft uses distributed estimation-based scheduling to scale to very large (\textgreater 20,000 node) clusters. Unlike Sparrow, which only takes into account machine queue depth, Apollo also takes into account load and data locality. Boutin and his colleagues were not as concerned with fairness, and they showed task scheduling rates over 10,000 per second. 

Microsoft's Mercury scheduler~[\cite{karanasos2015mercury}] is a hybrid (centralized and distributed) scheduler that allows applications to make trade-offs between the scheduling overhead centralized scheduling and the execution guarantees of distributed scheduling on very large clusters. Mercury was implemented as an extension to Apache Hadoop YARN. Rasler and his colleagues then extended and further refined the work of Sparrow (a distributed research scheduler mentioned below), Apollo, and Mercury with the Yaq-c (centralized) and Yaq-d (distributed) scheduling algorithms with more principled queue management techniques, including queue sizing, worker-side queue reordering, and placement of tasks.

\subsubsection{Open-Source Big Data Family}

Apache Hadoop YARN is an in-memory map-reduce job scheduler that enables scaling map-reduce style job environments to execute efficiently on several thousand servers. It was built to increase the processing speed and scalability of Apache Hadoop MapReduce [\cite{dittrich2012efficient}]. It is a monolithic scheduler with a simple API for batch map-reduce jobs, and it does not support microjobs or persistent jobs well. Several YARN add-ons introduce features into the scheduler. These add-ons include YARN Fair Scheduler~[\cite{bhattacharya2013hierarchical}], which enables fair job scheduling; Project Myriad, which enables YARN to run on a Mesos-managed cluster; Apache Llama (Low Latency Application Master), which enables batch and low-latency jobs via a bridge for YARN; and Apache TEZ, which enables job dependencies and directed, acyclic graph (DAG) dependencies in YARN.  

Apache Mesos is a two-level scheduler that enables the partitioning of a pool of compute nodes among many scheduling domains. Each scheduling domain has a pluggable scheduler called a Mesos framework (think queues and node pools), and each framework allocates the resources within its domain resources that have been allocated to it by Mesos. It has a rich API for communicating with the scheduler. Frameworks are written to use the Mesos API, thus making application integration on a Mesos cluster straightforward. 

Kubernetes is an open-source distributed Docker container management system that includes a scheduler. It is based on Google Borg and Omega, and it was released as an open-source project in 2015. The scheduler uses a simple first-in, first-out scheme; it has features to launch containers in an on-demand fashion; and it can queue up pending resource requests.

\subsubsection{Research Family}

Pacora [\cite{bird2014optimizing}, \cite{bird2011pacora}] is a research scheduler developed by Sarah Bird at UC Berkeley for her PhD thesis. She demonstrated the use of dynamic runtime resource utilization of each of the simultaneously executing application to better schedule and load-balance processes onto multicore computers. As future work, she considered extending these concepts to cluster scheduling by feeding back the dynamic state of resource utilization to the job scheduler and using Pacora algorithms to determine job placement more dynamically. Pacora implements resource allocation as a convex optimization. 

Among other research schedulers, the focus has been on data-dependent tasks; in other words, they do not schedule synchronously parallel job. A number of these schedulers explored the friction between scalability, placement correctness, and efficiency (including latency), particularly using distributed scheduling architectures. 

The Quincy scheduler~[\cite{isard2009quincy}] from Microsoft Research was the first distributed scheduler that tried to optimize the trade-offs between job data locality and fairness using a min-cost flow algorithm. It was purely a research project exploring distributed scheduling, and it is not available. 

Sparrow [\cite{ousterhout2013sparrow}] is a research scheduler developed at UC Berkeley that implements decentralized resource management and job scheduling. Sparrow is available as a Spark plug-in with no support. 

The Tarcil scheduler~[\cite{delimitrou2015tarcil}] developed by Stanford University and MIT is also a distributed scheduler that balances scheduling speed and resource matching  quality by employing dynamic job and resource statistics along with dynamic sub-sampling. 

Most recently, the Firmament centralized scheduler~[\cite{gog2016firmament}] improves on the min-cost max-flow (MCMF) scheduling optimization of the Quincy scheduler by using relaxation which has much lower scheduling latency.

\subsection{Scheduler Features}
Certain resource manager and scheduler features are key and the primary intent of the large-scale computing system and the nature of the job policies. There are many features that we can compare among these schedulers, and the most important features are briefly explained in this subsection. The features are explained with their impact on both HPC and big data jobs. The features are organized into subsections which will correspond to feature comparison subsections in Section~\ref{subsec:featurecomparison}.

\subsubsection{Scheduler Metadata}

\textit{Actively developed} assesses whether features are being added to the scheduler code base and if the code is being maintained. This is a bit of a double-edged sword. On the one hand, you want some development to occur, including adding some new features and porting to new operating systems and hardware. But if a project is too active, it can be difficult to keep up with the versions to deliver new features to users in a timely manner. 

\textit{Cost and licensing} conveys whether the scheduler is open source or has a license for which one must pay. While it is nice to have contracted support, it can be very expensive for large clusters. Free is not necessarily free, though. Even if a scheduler is open source and free, it is good to have in-house talent who are responsible for configuring and managing the scheduler. However, some licenses can be simply exorbitant with hefty per-node or even per-core charges.

\textit{Operating system support} captures the operating systems on which the scheduler runs and on which jobs can be executed. This feature is reasonably straightforward. While all of these schedulers run on Linux, distribution support may not be entirely universal (though you can usually compile the source code of open-source schedulers and possibly incur some porting overhead). Commercial schedulers may have a more limited list of supported Linux distributions, which are usually the more stable, commercially vetted ones like Red Hat Enterprise Linux, CentOS, and Suse, but perhaps not Fedora or Ubuntu. 

\textit{Language constraints} refers to the programming languages in which executed applications can be written. This feature is particularly important because data analysis applications are often written in rapid prototyping/scripting languages such as Matlab, Python, Java, Scala, and R, which are not the typical HPC programming languages.  Most of these schedulers can execute jobs in all of the major programming languages. However, with big data scheduler APIs, some languages are favored. Because many data analysis applications are often written with API calls to the scheduler, language support can be very important to data analysis applications. 

\textit{Access control and security} involves user authentication and isolation. Among the HPC schedulers, access control and security have always been parts of the feature set. Early versions of open-source big data schedulers did not have much security or isolation, but that is changing because of greater adoption and scaling to many more users.

\subsubsection{Job Types}

\textit{Job types} captures the parallel nature and size of the jobs. Among the number of ways to categorize jobs, the primary ways are in the dimensions of time executed (short, medium, long, very long) and the jobs' parallelism or lack thereof. This feature captures the length of jobs that the scheduler is designed to best accommodate. Many data analysis applications and services run as either fast-running small tasks or as persistently executing, long-running service jobs. Schedulers for data analysis need to be able to accommodate both of these job types well.

\textit{Parallel and job array support} captures the types of parallelism that the scheduler can handle. More specifically, there are single-process jobs; job arrays, in which multiple independent processes are run using a single job identifier with different parameters for each process; and parallel jobs, in which each of the processes are launched simultaneously and communicate during the computation. All schedulers handle single-process jobs. Among the HPC schedulers, all have support for both parallel jobs and job arrays, although in the PBS family of schedulers, the job array commands on the command line are not as well integrated as they are in the other HPC schedulers. Big data schedulers can run job arrays well, and it is possible to run parallel jobs though the synchrony in launching may not be robust.  

\textit{Parallel and array jobs} indicates whether the job scheduler allows synchronous dependent parallel and/or asynchronously independent parallel (array) jobs. 

\textit{Queue support} captures whether the scheduler can manage jobs with different characteristics and resource requirements in separate data structures called queues. Having multiple queues often makes it easier to manage jobs with disparately different requirements; however, having too many queues can make the user experience more confusing than it needs to be by presenting far too many choices of queues into which the users can submit their jobs. 

\textit{Metascheduling} is a feature that enable support for multiple resource managers. In the big data world, the challenge of executing long-running services alongside short-running analytics tasks can cause issues with their schedulers. The solution that the community settled on was multitiered scheduling in which a metascheduler chose which of one or more schedulers was best for each submitted job. This solution is not that different from the concept of queues among the HPC schedulers. We should note that  a large push for metascheduling in the supercomputing research community in the late 1990s and early 2000s enabled supercomputing jobs to be submitted and executed across multiple organizations and supercomputing centers including Globus [\cite{foster2003grid}, \cite{nabrzyski2004grid}], Legion [\cite{grimshaw1997legion}], and Condor Flocking [\cite{epema1996worldwide}]. That research has reached maturity and is being used by many supercomputing consortiums. However, these technologies are not being considered in this work because we are addressing data analysis systems that are not deployed across multiple organizations/centers.  

To a large extent, the queue support and metascheduler features attempt to solve the same problem from two different perspectives. HPC schedulers added queue support so that different types of jobs, resource policies, priorities, etc., were lumped together to help with prioritization, resource allocation, etc. Each queue can be assigned a certain set of compute nodes (and perhaps other resources) on which its jobs are allowed to run. The metascheduling feature in Mesos tries to solve a very similar problem by enabling users to submit their jobs to Mesos using a single common API.

\subsubsection{Job Scheduling}

\textit{Timesharing} is the ability to allocate multiple jobs from one or more users to a single compute node. Generally, since modern operating systems have that capability, so do the schedulers that feed jobs to them.

\textit{Backfilling} is the ability to schedule pending jobs when an executing job finishes early. Usually these backfilled jobs are smaller, both in execution time and number of job slots, than the larger parallel jobs. This is a feature that HPC schedulers have, while big data schedulers tend to schedule jobs on demand. 

\textit{Job chunking} has some relationship with job arrays, but is different. With job chunking, the scheduler searches through a given user's queued jobs to find similar jobs. It then schedules multiple similar jobs simultaneously. 

\textit{Bin packing} is a resource-allocation technique by which the scheduler chooses groups of jobs to launch simultaneously on a node or set of nodes to best utilize the node resources. 

\textit{Gang scheduling} allows a user to submit multiple processes to occupy a single job slot. Only one of the multiple processes can execute at any given time, but they are usually the type of processes that execute sequentially or only part of the time. 

\textit{Job dependencies and directed acyclic graphs (DAGs) support} is a feature that allows users to define execution dependencies between jobs. In data analysis applications, it is common to chain together analytics jobs with dependencies on each other, so support for this feature is essential.

\subsubsection{Resource Management}

\textit{Resource heterogeneity} refers to the scheduler's ability to accommodate differently configured nodes; i.e., they have different configurations of RAM, CPU, disk space, GPU availability, etc. More sophisticated resource heterogeneity allows administrator-defined resources that can be associated with compute nodes. 

\textit{Resource allocation policy} refers to the ability to write rules about how resources are utilized. Because short-running data analysis applications and long-running services usually require very different resource-usage profiles, it is important that these resource needs can be characterized precisely so that service jobs are provided only the resources that they really need since the allocated resources will be utilized by these long-running services for a long timeframe. 

\textit{Static and dynamic (consumable) resources} refers to how resources are consumed when a job is scheduled to use them. For instance, the total RAM on a compute node, as well as the CPU cores on a node, are static resources. Conversely, dynamic resources are resources that do change, such as load status and host status. 

\textit{Network-aware scheduling} takes the network topology into account when allocating jobs to nodes. The intent is to minimize network distance between jobs that need to communicate with each other in parallel jobs.

\subsubsection{Job Placement}

\textit{Intelligent scheduling} captures the fact that the scheduler is using a scheduling policy more sophisticated than first-in, first-out (FIFO). 

\textit{Prioritization schema}  assure that resources are allocated to jobs according to priorities. This function can be as simple as submitting jobs with a priority rating, but it can be much more sophisticated, such as having user priorities, user group priorities, and even fairshare policies implemented in the scheduler. 

\textit{Job replacement and reordering} refers to a feature by which a user can replace jobs and/or reorder the execution of jobs that are waiting to be executed in the job queue. 

\textit{Advanced reservations} is a feature that allows a user to request a set of job slots in the future. This feature is often used to enable interactive jobs on systems that use a highly utilized batch queue. 

\textit{Power-aware scheduling} works to minimize the number of nodes that are online and powered up so as to minimize the power being drawn by the system. This function not only involves packing jobs onto nodes effectively but also requires the ability to shut off and reboot nodes when necessary. 

\textit{User-related job placement} enables users to specify whether a job is scheduled (or not) on nodes on which another user's jobs are executing. 

\textit{Job-related job placement} refers to the ability to schedule jobs (or not) where other jobs are executing. 

\textit{Data-related job placement} schedules jobs on nodes on which the required data reside in local storage.

\subsubsection{Scheduling Performance}

\textit{Centralized vs. distributed scheduling} refers to the architecture with which job scheduling is accomplished. Schedulers need to keep track of jobs that have been submitted and are waiting for execution in a queue and jobs that have been dispatched and are currently executing. Each job has a data structure associated with it that captures a great deal of job information. Hence, most schedulers are centralized so that all of the job information is contained in that central location. In the past few decades, hot backup capabilities have been added to gain higher availability of the scheduler. However, a distributed scheduler architecture would allow for greater resilience but could cost the scheduler in performance. The Mesos scheduler is more distributed than the other schedulers in that it is distributed between its metascheduler role and each of the schedulers that it uses. 

\textit{Scheduler fault tolerance} is related to the centralized or distributed architecture of the scheduler, and it is usually enabled through a hot backup server or a set of distributed servers. It captures whether the scheduler and its jobs continue to execute despite a fault in the scheduler. 

\textit{Scalability and throughput} refer to how many jobs and nodes the scheduler can simultaneously schedule. This performance metric is particularly important to some big data parallel frameworks (particularly MapReduce) in which programmers are encouraged to write their pleasantly parallel software as rather short-duration, simple operations. 

\textit{Performance and bottlenecks} capture what the primary efficiency inhibitor is in scheduling jobs. Because of the complexity of receiving the status of node resources, managing the queue of pending jobs, performing resource allocations to available resources, dispatching jobs to nodes, and monitoring jobs in execution, it can be quite challenging to pinpoint bottlenecks. Further, the aforementioned list of functions does not even take into account the specific hardware and configurations of the system. 

\textit{Scheduler latency} refers to how quickly a job is dispatched onto job slots/resources once they become available.

\subsubsection{Job Execution}

\textit{Prolog/epilog support} is a feature that allows scripts to be executed before and/or after job execution. This functionality may include setting up certain execution environments or copying data. We have found this feature quite valuable for copying database datasets and virtual machine images onto and off of the local node storage in dynamic database and virtual machine capabilities. 

\textit{Data movement/file staging} is a built-in scheduler feature that enables the copying of supporting files onto the local storage of nodes. This feature was particularly important in the past when central file systems were less common. Most HPC schedulers have this feature, while big data schedulers generally do not. 

\textit{Checkpointing} allows applications to save the execution state of running jobs so that if a crash occurs, the job can be restarted from the checkpointed state. Most big data schedulers do not handle checkpointing but instead relaunch the job if it fails. 

\textit{Job migration} means that the scheduler can move a job from one compute resource to another while it is executing. Usually this feature requires scheduler-based checkpointing to record the current job state so that the job can be restarted from where it had left off on another node. This feature can be particularly useful in rebalancing long-running or service-oriented jobs. 

\textit{Job restarting} is a feature that restarts jobs if the job is aborted or fails (if so requested by the submitter). 

\textit{Job preemption} allows lower-priority executing jobs to be hibernated in order to execute higher-priority jobs. Usually this action involves using swap space on the compute nodes to hibernate the jobs.

\subsection{Feature Comparison}
\label{sub:featurecomparison}
Now that we have covered all of the scheduler features, we shall compare how the schedulers stack up against each other on the basis of these features. There are many overlaps and feature parity among the schedulers within each of the families so we will explore a subset of these schedulers. The schedulers must be openly available and have reasonable support and feature development through companies or consulting firms so that medium to large organizations have access to such outside mechanisms (except for the representative scheduler from the research scheduler family). We chose representative schedulers from each scheduler family. We have chosen the following set of schedulers for further evaluation:
\begin{itemize}
\item IBM Platform LSF
\item OpenLava
\item Slurm
\item Grid Engine
\item Pacora
\item Apache Hadoop YARN
\item Apache Mesos
\item Google Kubernetes
\end{itemize}
We will compare the representative schedulers on the features listed above. Please note that since Pacora is a research scheduler, many of the feature comparisons are not available for it. 

\subsubsection{Metadata Features Comparison}

\begin{table}
  \centering
  \caption{Metadata features comparison among job schedulers.}
  \label{tab:MetadataComparison}
  \scriptsize
  \begin{tabu} to \textwidth { | X[2l] || X[c] | X[c] | X[c] | X[c] | X[c] | X[c] | X[c] | X[c] |}
\hline
Metadata & \rot[90]{LSF} & \rot[90]{OpenLAVA} & \rot[90]{Slurm} & \rot[90]{Grid Engine} & \rot[90]{Pacora} & \rot[90]{YARN} & \rot[90]{Mesos} & \rot[90]{Kubernetes} \\ \hline \hline
    Type & HPC & HPC & HPC & HPC & HPC & Big Data & Big Data & Big Data \\ \hline
    Actively developed & \checkmark & \checkmark & \checkmark & \checkmark & -$^1$ & \checkmark & \checkmark & \checkmark \\ \hline
    Cost/ Licensing & \$\$\$ & Open source & Open source & \$\$\$, Open source & N/A & Open source & Open source & Open source \\  \hline
    OS support & Linux & Linux, Cygwin & Linux, *nix & Linux, *nix & N/A & Linux & Linux & Linux \\ \hline
    Language support & All & All & All & All & N/A & Java, Python$^2$ & All & All \\ \hline
    Access control/ security & \checkmark & \checkmark & \checkmark & \checkmark & - & \checkmark & \checkmark & \checkmark \\ \hline
    \hline
  \end{tabu}
  \vspace{1ex}

     \raggedright{$^1$Some further development is ongoing within Microsoft.                    
 $^2$Other languages are supported; Java and Python are the most strongly supported.}

  \end{table}

Table~\ref{tab:MetadataComparison} is a comparison of the metadata associated with the representative schedulers. All of the representative schedulers are being actively developed except for the research scheduler, Pacora, which has some further development within Microsoft. While LSF and Grid Engine each have a commercial version, the rest of the schedulers are available as open-source projects. Consulting and support contracts are available for them from various companies. As mentioned in Section~\ref{subsubsec:TraditionalHPCFamily}, OpenLAVA is an open-source version of LSF with reasonable feature parity, although the two schedulers are reported to not share a common code base. Slurm, LSF, and Grid Engine have been ported to many different operating systems on many different hardware platforms and network topologies over the years. All the production schedulers are supported on all the major server-based Linux distributions. 
Mesos is written primarily in C++ and should be portable to diverse (non-Linux) environments, if needed. 
Broad language support is almost universal except for YARN. YARN does support other languages to some extent, but Java and Python are the most strongly supported. 
While all of the schedulers have some access control and security, there is plenty of online discussion about how YARN, Mesos, and Kubernetes probably should have more. Each of the developer community project teams are addressing these concerns. 

\subsubsection{Job Type Features Comparison}

\begin{table}
  \caption{Job type features comparison among job schedulers.}
  \label{tab:JobTypeComparison}
  \scriptsize
  \begin{tabu} to \textwidth { | X[2l] || X[c] | X[c] | X[c] | X[c] | X[c] | X[c] | X[c] | X[c] |}
\hline
Job Types & \rot[90]{LSF} & \rot[90]{OpenLAVA} & \rot[90]{Slurm} & \rot[90]{Grid Engine} & \rot[90]{Pacora} & \rot[90]{YARN} & \rot[90]{Mesos} & \rot[90]{Kubernetes} \\ \hline \hline
    Parallel and array jobs & Both & Both & Both & Both & N/A & Array & Array & Array \\ \hline
    Queue support & \checkmark & \checkmark & \checkmark & \checkmark & - & \checkmark & \checkmark &  \\ \hline
    Multiple resource managers &  &  &  &  & - &  & \checkmark  &  \\ \hline
    \hline
  \end{tabu}
  \end{table}

The comparison of job type features is presented in Table~\ref{tab:JobTypeComparison}. The HPC schedulers have parallel jobs and array jobs, while the big data schedulers support array jobs. YARN, Mesos, and Kubernetes do not have built-in support for tightly coupled parallel jobs. Using the MPI framework with Mesos (or building a custom framework specifically for handling Slurm, PBS, Grid Engine, or LSF schedulers), Mesos can execute parallel jobs. 
Queue support is reasonably universal, but support for queues in big data schedulers is generally not as extensive as in the HPC schedulers. In YARN, queues are available in the capacity scheduler. In Mesos, one can think of each of the installed frameworks as a queue for different applications and/or application types. Each of the Mesos frameworks can implement its own set of queues to adjudicate between the jobs of its application flavor. Since the focus of Kubernetes is on container management, the scheduler has no queue support. 
Mesos is a metascheduler by design, and it allocates shared resources to other schedulers (frameworks). The other schedulers do not support multiple resource managers.

\subsubsection{Job Scheduling Features Comparison}

\begin{table}
  \caption{Job scheduling features comparison among job schedulers.}
  \label{tab:JobSchedulingComparison}
   \scriptsize
  \begin{tabu} to \textwidth { | X[2l] || X[c] | X[c] | X[c] | X[c] | X[c] | X[c] | X[c] | X[c] |}
\hline
Job Scheduling & \rot[90]{LSF} & \rot[90]{OpenLAVA} & \rot[90]{Slurm} & \rot[90]{Grid Engine} & \rot[90]{Pacora} & \rot[90]{YARN} & \rot[90]{Mesos} & \rot[90]{Kubernetes} \\ \hline \hline
    Timesharing & \checkmark & \checkmark & \checkmark & \checkmark & - & \checkmark & \checkmark & \checkmark \\ \hline
    Backfilling & \checkmark & \checkmark & \checkmark & \checkmark & - &  & - & - \\ \hline
    Job chunking &  &  &  & \checkmark & - &  & - & - \\ \hline
    Bin packing &  &  & \checkmark &  & - &  & - & - \\ \hline
    Gang scheduling &  &  & \checkmark &  & - &  & - & - \\ \hline
    Job dependencies and DAGs & \checkmark & \checkmark & \checkmark & \checkmark & - & \checkmark & -$^1$ &  \\ \hline
    \hline
  \end{tabu}
  \vspace{1ex}

     \raggedright{$^1$Job dependencies and DAG support in Mesos depends on having the feature in the plugged-in scheduler framework(s).}
  \end{table}

The job scheduling features are compared in Table~\ref{tab:JobSchedulingComparison}. Timesharing is critical for schedulers; one could easily argue that if a scheduler cannot handle scheduling multiple jobs to a compute node, it does not have a full feature set. Other features like backfilling, job chunking, bin packing, and gang scheduling are essential when systems have a very deep set of pending jobs in queues and there are expectations for the systems to be achieving 90\% or higher utilization. 
Similar to timesharing, job dependencies are also  essential for job schedulers. All of the HPC schedulers allow job dependencies to be submitted. Mesos can support job dependencies if a given plugged-in framework supports job dependencies. 
Kubernetes does not support dependencies; in its container-based environment, it is expected that users handle job dependencies at the application level. 

\subsubsection{Resource Management Features Comparison}

\begin{table}
  \caption{Resource management features comparison among job schedulers.}
  \label{tab:ResourceManagementComparison}
  \scriptsize
  \begin{tabu} to \textwidth { | X[2l] || X[c] | X[c] | X[c] | X[c] | X[c] | X[c] | X[c] | X[c] |}
\hline
Resource Management & \rot[90]{LSF} & \rot[90]{OpenLAVA} & \rot[90]{Slurm} & \rot[90]{Grid Engine} & \rot[90]{Pacora} & \rot[90]{YARN} & \rot[90]{Mesos} & \rot[90]{Kubernetes} \\ \hline \hline
    Resource heterogeneity & \checkmark & \checkmark & \checkmark & \checkmark & \checkmark & \checkmark & \checkmark & \checkmark \\ \hline
    Resource allocation policy & \checkmark & \checkmark & \checkmark & \checkmark & \checkmark & \checkmark & \checkmark & \checkmark \\ \hline
    Static and dynamic resources & Both & Both & Both & Both & Both & Both & Both & Both \\ \hline
    Network aware scheduling & \checkmark &  & \checkmark & \checkmark & - &  &  &  \\ \hline
    \hline
   \end{tabu}
 \end{table}

Table~\ref{tab:ResourceManagementComparison} compares resource management features among the representative job schedulers. Similar to timesharing and job dependencies, resource management and resource allocation policies are essential to the functionality of a modern scheduler. However, how many types of resources a scheduler offers (including site-defined resources) and how optimally jobs are mapped onto the resources is  NP-complete and continues to be an open research problem. All of the schedulers support policy-based resource allocation and static and dynamic resources. Static resources are either busy or not (i.e., a job slot), while dynamic resources can be partially used by jobs (i.e., the memory of a compute node). Further, all the schedulers have mechanisms for administrators to set and manage resources via configuration scripts. 
In YARN, there is some concept of network-aware scheduling in that jobs are dispatched to different parts of the dataset that are stored on each of the Hadoop nodes. Usually files in the Hadoop distributed file system (HDFS) are replicated three times by default. There is the original file, a replica of the file somewhere in the same rack of the data center, and another replica in a different rack of the data center. This distribution scheme is intended to increase the redundancy of each file to provide resilience to failures. However, the data location-oriented job scheduling in Hadoop YARN is not nearly as sophisticated as the network-aware scheduling that is available in LSF,  Slurm, and Grid Engine (as well as PBS).

\subsubsection{Job Placement Features Comparison}

\begin{table}
  \caption{Job placement features comparison among job schedulers.}
  \label{tab:JobPlacementComparison}
  \scriptsize
  \begin{tabu} to \textwidth { | X[2l] || X[c] | X[c] | X[c] | X[c] | X[c] | X[c] | X[c] | X[c] |}
\hline
Job Placement & \rot[90]{LSF} & \rot[90]{OpenLAVA} & \rot[90]{Slurm} & \rot[90]{Grid Engine} & \rot[90]{Pacora} & \rot[90]{YARN} & \rot[90]{Mesos} & \rot[90]{Kubernetes} \\ \hline \hline
    Intelligent scheduling & \checkmark & \checkmark & \checkmark & \checkmark & - & -$^1$ & \checkmark$^2$ &  \\ \hline
    Prioritization schema & \checkmark & \checkmark & \checkmark & \checkmark & - & \checkmark & \checkmark & \checkmark \\ \hline
    Job replacement and reordering & \checkmark & \checkmark & \checkmark & \checkmark & - &  & \checkmark & \checkmark \\ \hline
    Advanced reservations & \checkmark &  & \checkmark & \checkmark & - &  &  &  \\ \hline
    Power-aware scheduling & \checkmark &  & \checkmark & \checkmark & - &  &  &  \\ \hline
    User-related job scheduling & \checkmark &  & \checkmark & \checkmark & - &  &  &  \\ \hline
    Job-related job scheduling & \checkmark &  & \checkmark & \checkmark & - &  &  &  \\ \hline
    Data-related job scheduling &  &  &  &  & - & \checkmark &  &  \\ \hline
    \hline
  \end{tabu}
  \vspace{1ex}

     \raggedright{$^1$Both the Fair Scheduler and Capacity Scheduler in YARN offer some intelligent scheduling. 
 $^2$Intelligent scheduling depends on intelligent scheduling support in the plugged-in scheduler framework(s).}

  \end{table}
Job placement features are compared in Table~\ref{tab:JobPlacementComparison}. Each of the HPC schedulers and Mesos have available scheduling algorithms that are more sophisticated than first-in, first-out. YARN has more sophisticated scheduling algorithms in its Fair Scheduler and Capacity Scheduler, while Kubernetes does not include sophistication beyond first-in, first-out. 
Every scheduler enables some prioritization scheme though it may be primitive. Many of the other scheduling techniques (advanced reservations, power-aware scheduling, user-related job scheduling, and job-related job scheduling) are more applicable to batch-queue systems with long-running jobs. These techniques are reasonably complex so it is not unreasonable for purely open-source projects like OpenLAVA to not have such a feature. Finally, only YARN supports data-related job scheduling, but one can develop scripts that read location data from distributed file systems like Hadoop distributed file system [\cite{shvachko2010hadoop}] and annotate dynamically generated submission scripts to have similar effects.

\subsubsection{Scheduling Performance Features Comparison}

\begin{table}
  \caption{Scheduling performance features comparison among job schedulers.}
  \label{tab:SchedulingPerformanceComparison}
  \scriptsize
  \begin{tabu} to \textwidth { | X[2l] || X[c] | X[c] | X[c] | X[c] | X[c] | X[c] | X[c] | X[c] |}
\hline
Scheduling Performance & \rot[90]{LSF} & \rot[90]{OpenLAVA} & \rot[90]{Slurm} & \rot[90]{Grid Engine} & \rot[90]{Pacora} & \rot[90]{YARN} & \rot[90]{Mesos} & \rot[90]{Kubernetes} \\ \hline \hline
    Centralized vs. distributed & Cent. & Cent. & Cent. & Cent. & Cent. & Cent. & Dist. & Cent. \\ \hline
    Scheduler fault toleraance & \checkmark &  & \checkmark & \checkmark & - & \checkmark & \checkmark & \checkmark \\ \hline
    Scalability and throughput & 10K+ & 1K+ & 100K+ & 10K+ & - & 10K+ & 100K+ & 100K+ \\ \hline
    \hline
  \end{tabu}
  \end{table}

Table~\ref{tab:SchedulingPerformanceComparison} compares the scheduling performance of the representative schedulers. All of the schedulers except Mesos employ a centralized scheduler, and all but OpenLAVA and Pacora implement fault tolerance with one or more spares. Mesos is a distributed scheduler by its metascheduling nature; each of the framework schedulers are distributed. Each of the schedulers is able to handle anywhere from thousands to hundreds of thousands of simultaneously executing job slots. This number depends upon design and implementation of the job and resource data structures and management algorithms. 
OpenLAVA was just recently reinvigorated, so it would not be surprising for its open-source team to add fault tolerance and greater scalability soon. 
Performance, bottlenecks, and scheduler latency are addressed in Section~\ref{sec:Performance}.

\subsubsection{Job Execution Features Comparison}

\begin{table}
  \caption{Job execution features comparison among job schedulers.}
  \label{tab:JobExecutionComparison}
  \scriptsize
  \begin{tabu} to \textwidth { | X[2l] || X[c] | X[c] | X[c] | X[c] | X[c] | X[c] | X[c] | X[c] |}
\hline
Job Execution & \rot[90]{LSF} & \rot[90]{OpenLAVA} & \rot[90]{Slurm} & \rot[90]{Grid Engine} & \rot[90]{Pacora} & \rot[90]{YARN} & \rot[90]{Mesos} & \rot[90]{Kubernetes} \\ \hline \hline
    Prolog/epilog support & \checkmark &  & \checkmark & \checkmark & - &  & \checkmark & \checkmark \\ \hline
    Data movement/file staging & \checkmark &  & \checkmark & \checkmark & - &  &  &  \\ \hline
    Checkpointing & \checkmark & \checkmark & \checkmark & \checkmark & - &  &  &  \\ \hline
    Job migration & \checkmark & \checkmark & \checkmark & \checkmark & - &  & ? & ? \\ \hline
    Job restarting & \checkmark & \checkmark & \checkmark & \checkmark & - & \checkmark & \checkmark & \checkmark \\ \hline
    Job preemption & \checkmark & \checkmark & \checkmark & \checkmark & - &  & \checkmark & \checkmark \\ \hline
    \hline
  \end{tabu}
  \end{table}

Finally, Table~\ref{tab:JobExecutionComparison} compares job execution features among the representative job schedulers. 
Prolog and epilog scripting support is fairly well covered among the schedulers while data movement and file staging are supported by the more established HPC schedulers. The big data schedulers leave that functionality to be included in applications. Similarly, HPC schedulers include scheduler-based checkpointing while big data schedulers also leave that functionality to the applications. Data analysis applications need restarting, and migration is very helpful for job fault tolerance; restarting is well supported by the schedulers while migration of jobs is done by the programmers/users in Mesos and Kubernentes. Job preemption support in the scheduler is available in the HPC schedulers and in Mesos and Kubernetes. 

\subsection{Summary of Observations}
\label{subsec:SummaryOfObservations}

These comparisons show that there is a great amount of parity among features. Such features common across the majority of traditional HPC and big data schedulers include timesharing, queues, job dependencies, heterogeneous resources, static and dynamic resources, intelligent scheduling, prioritization schema, fault tolerance job restarting, and prolog/epilog support. The depth and complexity of how the features are implemented in each of the schedulers may vary greatly, and one can argue that such details usually only affect the power users of the system and have little to no impact on the novice and average users. There is another set of more complex features that are generally unique to traditional HPC workloads. These features include synchronously parallel job launches, advanced job scheduling features (backfilling, job chunking, bin packing, and gang scheduling),  network-aware scheduling, advanced job placement features, and checkpointing. 
Finally, big data schedulers stand out in having easy-to-use APIs with which applications are developed, widely shared and distributed, and executed, and that lower the barrier to adoption and utilization. However, the aforementioned DRMAA API is quite similar to these APIs, and it could be utilized to a similar effect. 

In the scheduling performance subsection, we postponed addressing performance, bottlenecks, and scheduler latency. These are the key subjects in the next sections.

\section{Latency and Utilization Modeling}
\label{sec:Modeling}

\begin{table}
  \centering
  \caption{Summary of notation.}
  \label{tab:NotationSummary}
  \begin{tabular}{ | l || l | }
    \hline
    Notation & Definition \\ \hline \hline
	$t_s$ & Scheduler latency \\ \hline
	$t$ & Task time \\ \hline
	$T_{total}$ & Total runtime of a job set \\ \hline
	$T_{job}$ & Total isolated job execution time \\ \hline
	$\Delta T$ & Non-execution latency \\ \hline
	$N$ & Total tasks \\ \hline
	$n$ & Tasks per processor  \\ \hline
	$\alpha_s$ &  Exponent accounting for nonlinear behavior in the scheduler \\ \hline
	$U$ & System utilization  \\ \hline
	$U_c$ & System utilization for constant time tasks \\ \hline
	$U_v$ & System utilization for variable time tasks \\ \hline
    \hline
  \end{tabular}
\end{table}

In order to better understand scheduler latency, it is worthwhile to construct a mathematical model for this latency. A summary of the notation used in this section is provided in Table~\ref{tab:NotationSummary}. Scheduler latency $t_s$ is the overhead associated with executing a task with the scheduler. The scheduler latency is influenced by submission, queue management, resource identification, resource selection, resource allocation, job dispatch, and job tear down. Task time $t$ is the time the task takes to run in complete isolation on a given resource, i.e., compute node. The value of $t$ then falls into one of two cases: 
\begin{itemize}
\item Case 1: Low scheduler latency impact -- $t \gg t_s$
\item Case 2: High scheduler latency impact -- $t \lessapprox t_s$
\end{itemize}
Data analysis jobs often execute many short-duration tasks, so they are generally more susceptible to realizing case 2. 

The diversity of steps that are performed by a scheduler makes latency difficult to measure. First, it is difficult to position a single observer in the system that will work well with all of the schedulers. Further, there are so many measurements that would need to be taken that the measurements can negatively affect the scheduler latency. However, all is not lost. The runtime model of the overall scheduler latency can be derived from the measured total runtime of a job $T_{total}$ consisting of $N$ tasks running on $P$ processors 
\begin{equation*}
    T_{total}(N,P) = T_{job}(N,P) + \Delta T(N,P)
\end{equation*}
where the isolated job execution time $T_{job}$ is the isolated task times on a processor, and the difference between the total runtime and the isolated job time is $\Delta T$. This equation then leads to the following latency model for each component on the right of the equation. If the job consists of tasks with the same runtime on every processor, then $T_{job}$ and $\Delta T$ can be modeled as 
\begin{align*}
    T_{job} = t n && \Delta T = t_s n^{\alpha_s}
\end{align*}
where $n = N/P$ is the number of tasks that are run on one processor, $t_s$ is the marginal scheduler latency incurred by adding a task to a processor, and $\alpha_s \approx 1$ is the exponent that accounts for nonlinear behavior in the scheduler. (Note, a number of equation formulations were considered for $\Delta T$, and it was this exponential equation that best fit all of the data that we collected and analyzed.) On the basis of the above model, there are three ways to reduce the impact of the scheduler launch latency on the total runtime: reduce $n$, $t_s$, and/or $\alpha_s$. 

The above model assumes that all tasks in a job have equivalent runtimes. However, we can account for different runtimes in the following way. If the job consists of tasks with different runtimes $t_j$, then 
\begin{align*}
    T_{job}(p) = \sum_{j \in J(p)} t_j && \Delta T = t_s n(p)^{\alpha_s}
\end{align*}
where $J(p)$ is the set of tasks that run on processor $p$, and $n(p)$ is the number of tasks that run on processor $p$. Variable task times could impact the scheduler performance if $T_{job}(p)$ and/or $n(p)$ are sufficiently different from their averages. 

Now that we have defined the scheduler latency model, we can establish the definition for utilization. Utilization $U$ is the time spent working on a task $T_{job}$ divided by the total time $T_{total}$ that the resource is claimed by the scheduler to execute the task: 
\begin{equation*}
    U = T_{job} / T_{total} 
\end{equation*}
This can be rewritten as
\begin{equation*}
	U^{-1} = 1 + \Delta T / T_{job}
\end{equation*}
For constant task times, utilization can be modeled as 
\begin{equation*}
	U_c^{-1} = 1 + (t_s n^{\alpha_s}) / (t n)
\end{equation*}
For $\alpha_s \approx 1$, the above formula simplifies to 
\begin{equation*}
	U_c(t)^{-1} \approx 1 + t_s / t
\end{equation*}
The above model illustrates the strong impact the scheduler latency can have on utilization for tasks similar in duration to $t_s$:
\begin{equation*}
	t_s \approx t  \Rightarrow  U_c^{-1} \approx 2  \Rightarrow  U_c \approx 0.5
\end{equation*}
For variable task times, utilization can be modeled then as
\begin{equation*}
	U_v(p)^{-1} = 1 + (t_s n(p)^{\alpha_s}) / (\sum_{j \in J(p)} t_j)
\end{equation*}
For $\alpha_s \approx 1$, this simplifies to 
\begin{equation*}
	U_v(p)^{-1} \approx 1 + (t_s n(p)) / (\sum_{j \in J(p)} t_j) 
					= 1 + t_s/t(p) = U_c(t(p))^{-1}
\end{equation*}
where $t(p)$ is the average time of the tasks that ran on processor $p$: 
\begin{equation*}
	t(p) = (\sum_{j \in J(p)} t_j) / n(p)
\end{equation*} 
If the scheduler releases a processor as it completes its work, then the overall utilization is the average of the per-processor utilization
\begin{equation*}
	U^{-1} \approx P^{-1} \sum_p U_c(t(p))^{-1}
\end{equation*}
By measuring the constant time task utilization $U_c(t)$ curve, it is possible to estimate the utilization for any distribution of tasks from the average task times on each processor. 

There are several takeaways that we can gain from this scheduler latency model. First, scheduler latency can have a large impact on data analytics workloads that have many short-duration tasks. There are three key parameters that impact scheduler performance: 
\begin{itemize}
	\item $n$, the number of tasks that ran on one processor,
	\item $t_s$, the marginal scheduler latency incurred by adding a task to a processor, 
		and
	\item $\alpha_s \approx 1$, the exponential that accounts for nonlinear behavior in 
		the scheduler. 
\end{itemize}
Scheduler performance can be improved by reducing the value of these parameters. Finally, the utilization curve for jobs with constant time tasks can be used to estimate the utilization of jobs with variable time tasks. In the next section, we will develop a benchmark with constant time tasks and evaluate the utilization and scaling of job scheduling latency with these constant time tasks. This approach allows us to characterize scheduler performance without having to develop more complex variable time task distributions.

\section{Performance Comparison}
\label{sec:Performance}
If batch scheduling is used, it means that there are consistently (or at least on average) more jobs to be executed than there are computational resources to execute them during a given time interval. With a system that is designed for batch processing large parallel jobs, it is important that the scheduler is able to schedule these large parallel jobs while interspersing shorter, smaller jobs, including job arrays, in order to keep resource utilization high. Thus, HPC schedulers need sophisticated job scheduling algorithms in order to utilize the computational resources most effectively. Conversely, on-demand scheduling is ideal for interactive data analysis jobs because the user does not have to wait for his/her job to percolate through the batch queue before interacting with it. If the system is used mainly to process on-demand interactive jobs such as data analysis jobs, the resource manager must quickly update (with timely periodicity) the status of all of the nodes so that the job scheduler is able to quickly assign new on-demand jobs to resources. This responsiveness is measured by scheduler latency, which involves the latency of submission, queue management, resource identification, resource selection, resource allocation, job dispatch, and job termination.

\begin{figure}[htb]
    \centering
    \includegraphics[width=30pc]{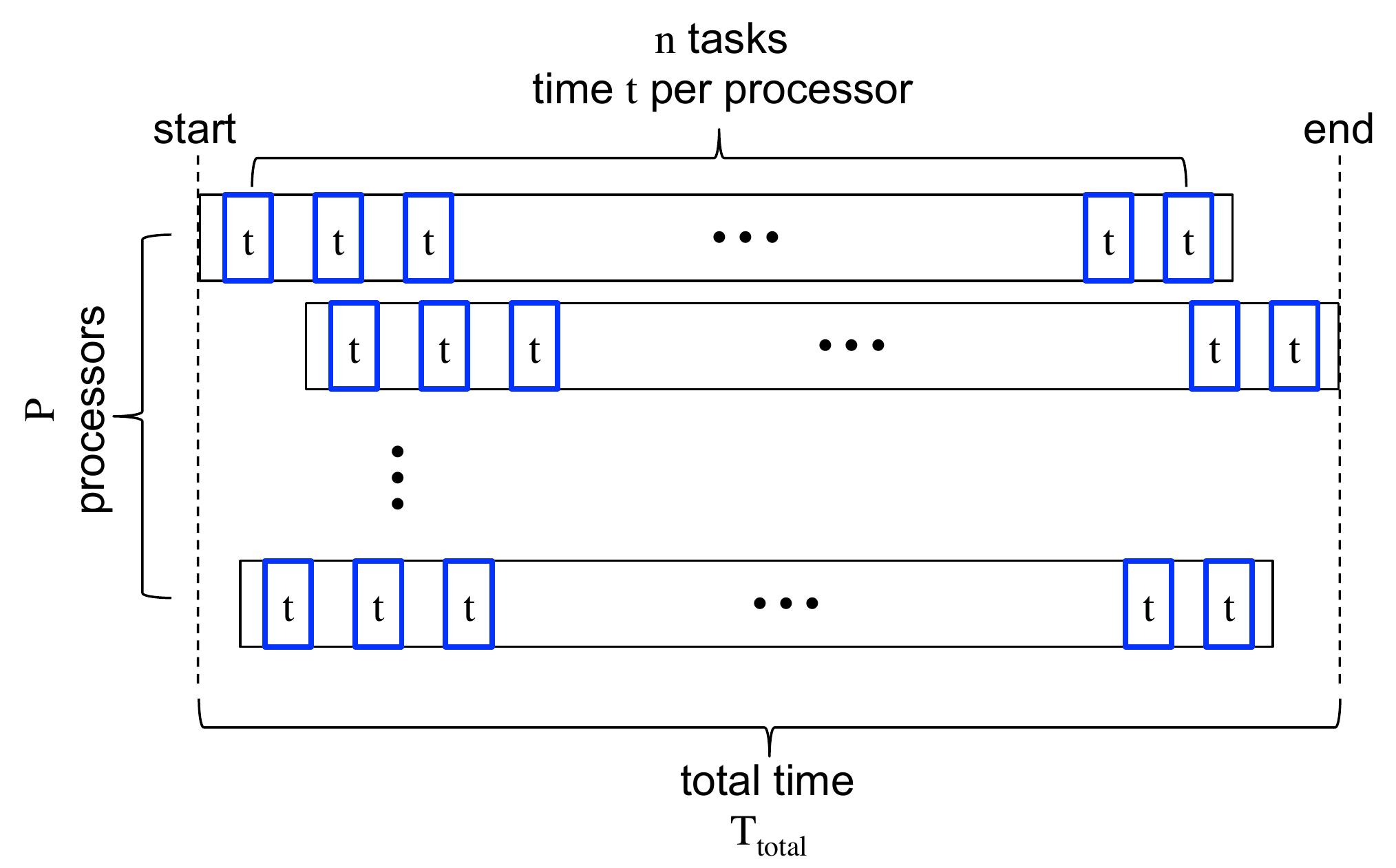}
    \caption{Schematic of the execution of a job $P$ processors, consisting of $n$ tasks each executing for time $t$. The total time, $T_{total}$, includes both the task execution time along with all scheduler overheads.}
    \label{fig:TestPlan}
  \end{figure}

Building on the model of Section~\ref{sec:Modeling}, the experiment test plan is depicted in Figure~\ref{fig:TestPlan}. We schedule $n$ constant time tasks such that each executes in $t$ time on $P$ processor cores. The isolated job execution time is then
\begin{equation*}
    T_{job} = t \cdot n 
\end{equation*}
We measure the total time to execute the $n \cdot P$ tasks as $T_{total}$. We can then calculate the scheduler utilization as 
\begin{equation*}
   U = T_{job} / T_{total} 
\end{equation*}
This test plan allows us to determine the scaling behavior across schedulers for a large number of short-duration jobs by producing curves of utilization $U$ versus task time. As we derived in Section~\ref{sec:Modeling}, by measuring the constant time task utilization $U_c(t)$, we can estimate the utilization for any distribution of tasks from the average task times on each processor. Hence, we can be confident that any combination of variable task times can be estimated using these constant task time results. 

One must keep in mind that we are evaluating the performance of the scheduler from the scheduler perspective. That is, we are executing constant time tasks to occupy a given job slot for a set amount of time. So for that job slot, the scheduler does not need to reevaluate its assignment until that occupation time has expired. The constant time tasks are inserting delays after which the scheduler must schedule and dispatch a new task to a given resource. There are naturally other considerations such as slowdown of task execution due to resource contention that affect the amount of time it takes to complete the task. However, such considerations are an issue of scheduling functionality effectiveness rather than scheduler latency. 

\subsection{Benchmarking Environment}
To minimize uncertainty, all measurements were carried out with our MIT SuperCloud cluster [\cite{reuther2013llsupercloud}].  This cluster was fully isolated from extraneous processes in order to deliver consistent results.  Forty-five nodes were used: one to serve as the scheduler node and forty-four as compute nodes (1408 cores total). All of the nodes were connected via a 10 GigE interface to a Lustre parallel storage array.

%

To compare the launch latency for the schedulers, we chose four representative schedulers from across the scheduler landscape: Slurm, Son of Grid Engine, Mesos, and Hadoop YARN. Each of the four schedulers was installed on the scheduler node and compute nodes, and the daemons of only the scheduler under test were running; i.e., only one scheduler was running at a time. Each of the schedulers had numerous parameters that needed be tuned for optimal performance on short tasks.   All four were configured to achieve optimal performance on short-duration jobs. 

The version designators and scheduler configuration details are as follows. We used Slurm version 15.08.6 with the following parameter settings:

{\tt ProctrackType = proctrack/cgroup}

{\tt SchedulerType = sched/builtin}

{\tt SelectType = select/cons\_res}

{\tt SelectTypeParameters = CR\_Core\_Memory}

{\tt PriorityType = priority/basic}

{\tt DefMemPerCPU = 2048}

\noindent For the Grid Engine testing, we used Son of Grid Engine 8.1.8 and enabled  the high-throughput configuration. With Mesos, we used Mesos version 0.25.0 with a single-node master and single ZooKeeper running on the master. Finally, with Apache Hadoop YARN, we used version 2.7.1 with a single-node master running NameNode and ResourceManager (YARN) daemons and compute nodes running DataNode and NodeManager daemons.

\subsection{Latency and Utilization Measurements}
\label{subsec:latencyandutlization}

\begin{table}
  \centering
  \caption{Parameter sets and runtimes used to measure scheduler latency as a function of job task time. Each task set was run three times for each scheduler, and all three times are listed.  Note: The Hadoop YARN Rapid Task trial set was not executed because execution time was exceedingly long.}
  \label{tab:parameters1}
  \begin{tabular}{ | l || c | c | c | c | }
    \hline
    \multirow{2}{*}{Configuration} & Rapid & Fast & Medium & Long \\
       & Tasks & Tasks & Tasks & Tasks \\ \hline \hline
Task time $t$ & 1 sec & 5 secs & 30 secs & 60 secs \\ \hline
Job time per processor $T_{job}$ & 240 secs & 240 secs & 240 secs & 240 secs \\ \hline
Tasks per processor $n$ & 240 & 48 & 8 & 4 \\ \hline
Processors $P$ (cores) & 1408 & 1408 & 1408 & 1408 \\ \hline
Total tasks $N$ & 337920 & 67584 & 11264 & 5632 \\ \hline
Total processor time & 93.7 hours & 93.7 hours & 93.7 hours & 93.7 hours \\ \hline
Runtimes (sec)  &  &  &  &  \\ \hline
  Slurm & 2774, 2787, 2790 & 622, 603, 606 & 280, 278, 255 & 287, 264, 300 \\ \hline
  GE & 3057, 3073, 3082 & 622, 634, 623 & 278, 279, 277 & 275, 281, 274 \\ \hline
  Mesos & 1794, 1795, 1792 & 366, 364, 367 & 280, 280, 281 & 306, 306, 305 \\ \hline
  Hadoop YARN & -- & 2013, 1798, 1710 & 479, 472, 510 & 342, 445, 347  \\ \hline
    \hline
  \end{tabular}
\end{table}

The jobs that were launched on the 1408 cores were all sleep jobs of 1, 5, 30, or 60 seconds. The total number of tasks $N$ and the number of tasks per processor $n$ were set so that the total processing time was always the same: 93.7 hours (337,920 seconds). The four task sets used for comparison are given  in Table~\ref{tab:parameters1}. For each task set and each scheduler, three trials were executed, and the results are the average of the three trials. The Hadoop YARN trials for rapid tasks were abandoned because it took too much time to execute because of excessive scheduler latency. With all four schedulers, the jobs were submitted as job arrays because they introduce much less scheduler latency than is introduced by submitting the same workload as individual jobs. 

Each of the task sets is composed of only one duration length of jobs. As we determined from the scheduler latency model of the previous section, heterogeneous mixes of job durations can be composed of combinations of these task sets (and other in-between duration values); hence, the trends that we learn from these task sets can easily be used to deduce the scheduler latency performance of more heterogeneous mixes of job durations.
  \begin{figure}[!ht]
    \subfloat[Slurm]{%
      \includegraphics[width=0.45\textwidth]{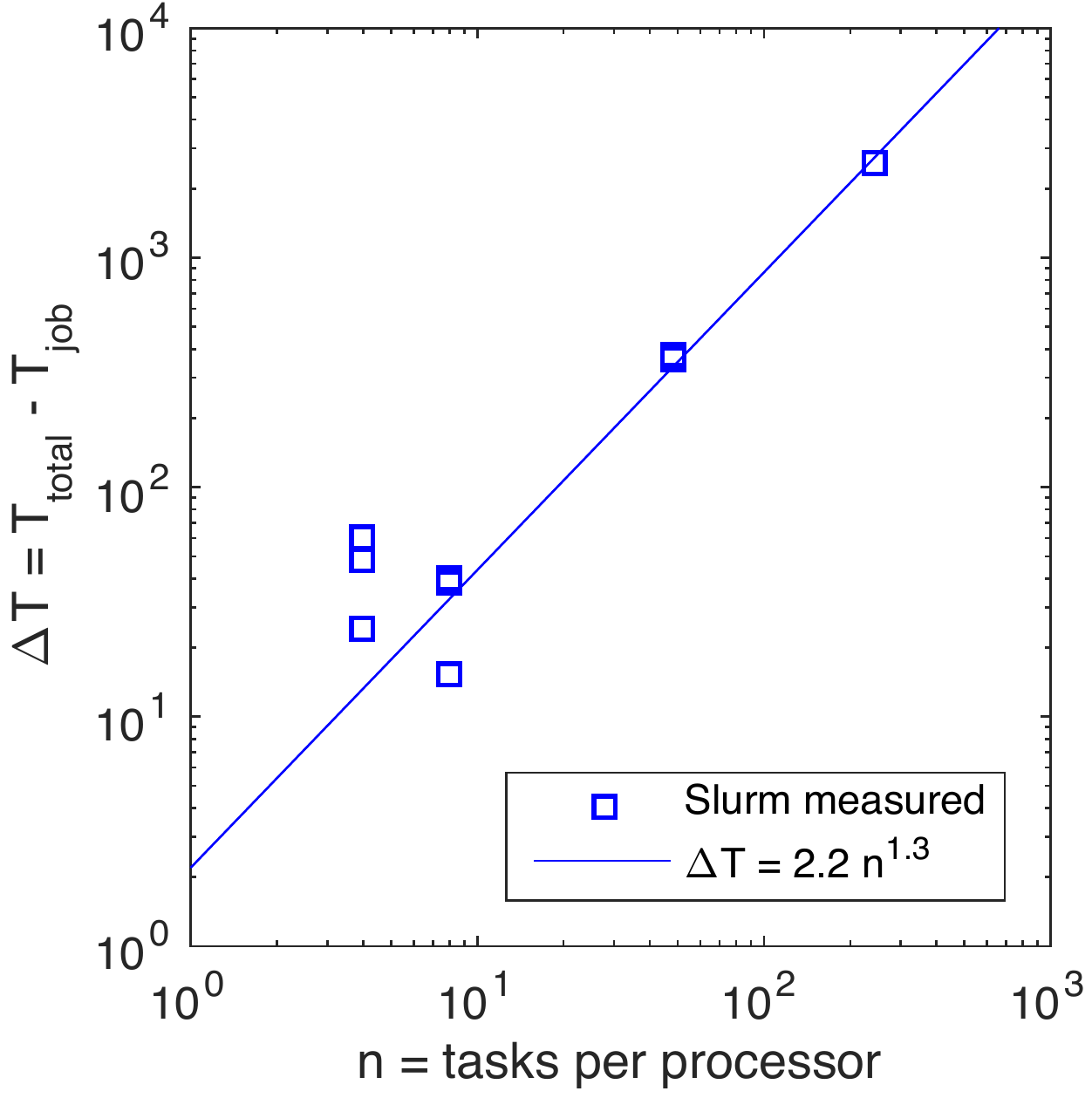}
    \label{subfig-1:LCa}
    }
    \hfill
    \subfloat[Grid Engine]{
      \includegraphics[width=0.45\textwidth]{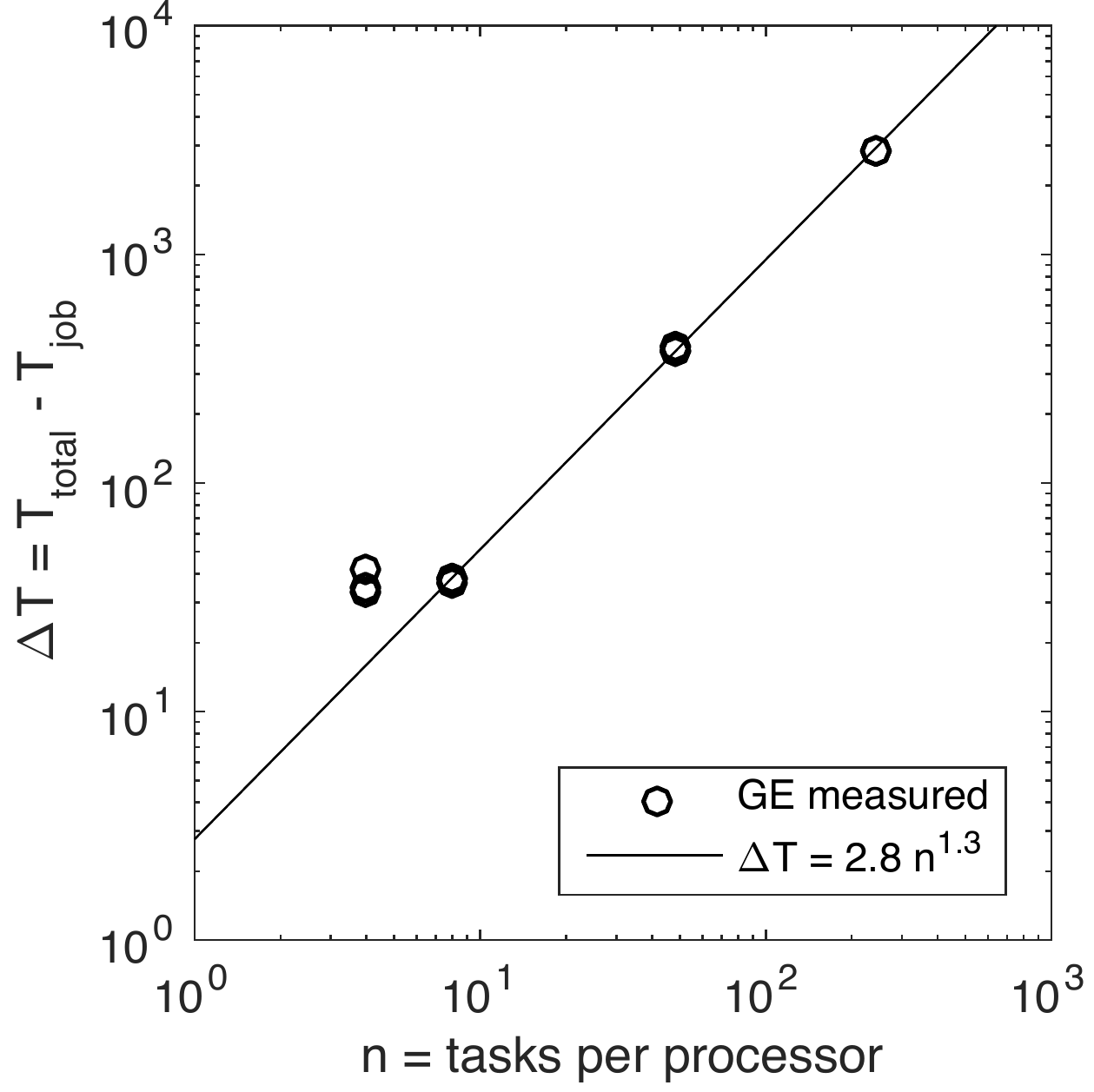}
    \label{subfig-2:LCb}
    } \\
    \subfloat[Mesos]{%
      \includegraphics[width=0.45\textwidth]{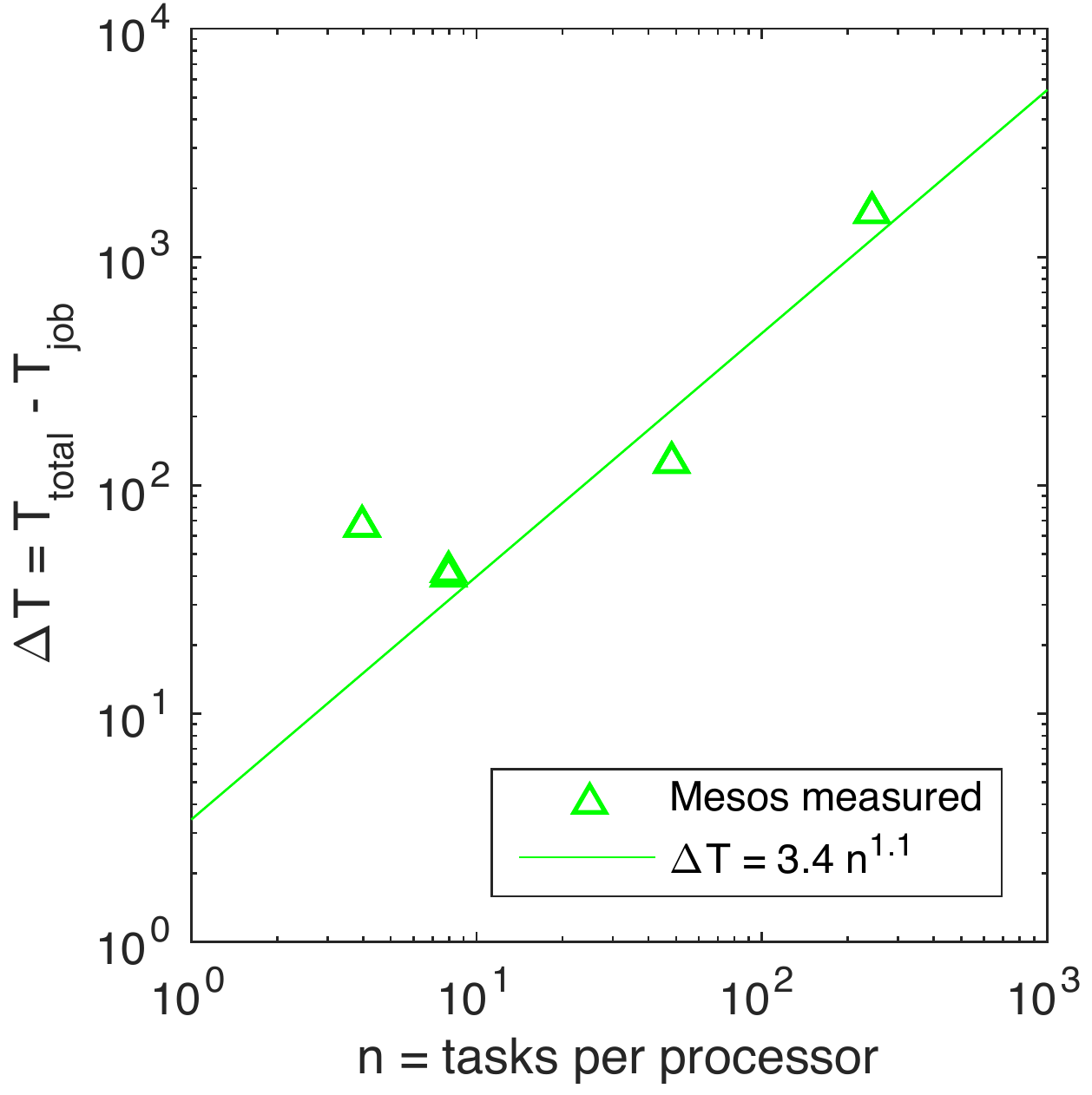}
    \label{subfig-3:LCc}
    }
    \hfill
   \subfloat[Hadoop YARN]{
      \includegraphics[width=0.45\textwidth]{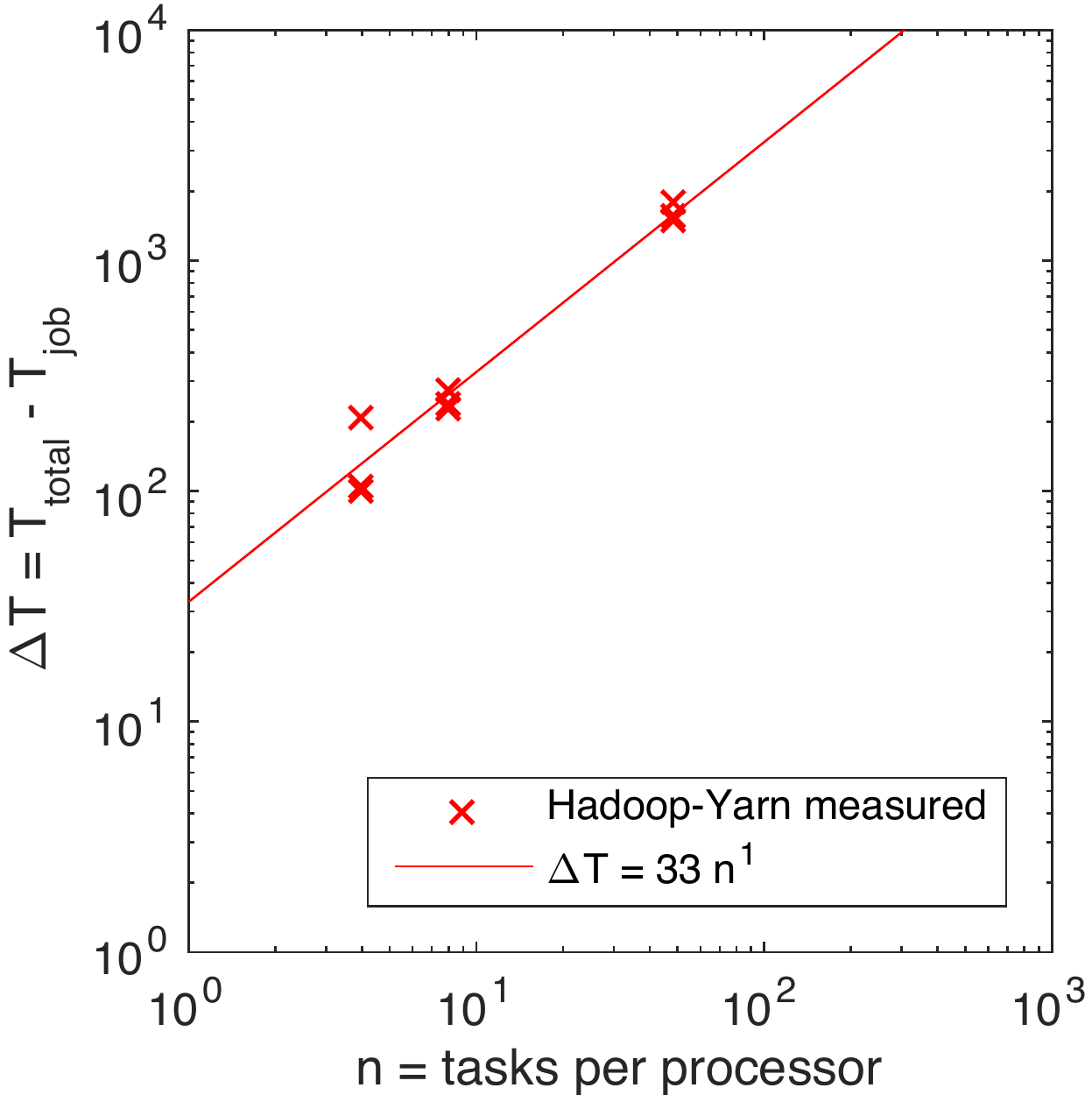}
    \label{subfig-4:LCd}
    }
   \caption{Comparison of scheduler latency versus number of tasks per processor.}
    \label{fig:LatencyComparison}
  \end{figure}

First, we analyzed the distribution of $\Delta T$ for each of the four schedulers under test as $n$, the number of tasks per processor, was increased. These results are shown in Figure~\ref{fig:LatencyComparison}. Note that these plots are on log-log axes.  Figure~\ref{subfig-1:LCa} shows that Slurm has low scatter among trials at higher values of $n$, and the model is a good fit at higher values of $n$. At lower values of $n$, shot noise impacts the model. For Grid Engine, there is low scatter among trials at all values of $n$, and the model is a good fit at most values of $n$, as seen in Figure~\ref{subfig-2:LCb}. Similar to Slurm, shot noise impacts the model at lower values of $n$. In Figure~\ref{subfig-3:LCc} for Mesos, there is very low scatter among trials at all values of $n$, and the model is a reasonable fit at most values of $n$, though there is some curvature away from the power law. There is some shot noise impact on the model at lower values of $n$. Finally, in Figure~\ref{subfig-4:LCd} for the Hadoop YARN scheduler, there is some scatter among trials at all values of $n$, and the model is a reasonable fit at all values of $n$. At the highest values of $n$, large marginal latency limits the ability to conduct experiments at all. 

\begin{table}
  \centering
  \caption{Measured model fit parameter comparison for scheduler latency model.}
  \label{tab:LatencyParameterComparison}
  \begin{tabular}{ | l || c | c | }
    \hline
    \multirow{2}{*}{Scheduler} & Marginal Latency $t_s$ & Nonlinear Exponent $\alpha_s$ \\
       & (smaller is better) & (smaller is better) \\ \hline \hline
	Slurm & 2.2 seconds & 1.3 \\ \hline
	Grid Engine & 2.8 seconds & 1.3 \\ \hline
	Mesos & 3.4 seconds & 1.1 \\ \hline
	Hadoop YARN & 33 seconds & 1.0 \\ \hline
    \hline
  \end{tabular}
\end{table}

It is also interesting to examine the model fit parameters for each of the schedulers. The model fit parameters are shown in Table~\ref{tab:LatencyParameterComparison}, where the second column is the $y$-axis crossing points and the third column is the angle of the fit line in the log-log plot. For this set of experiments, Slurm has the best marginal latency, while Grid Engine and Mesos have acceptable marginal latency. This result suggests that these three schedulers do not use excessive overhead to launch jobs. Mesos and Hadoop YARN have the best nonlinear exponent, suggesting that they have more uniform per-task launch latency overhead. 

\begin{figure}[htb]
    \centering
    \includegraphics[width=30pc]{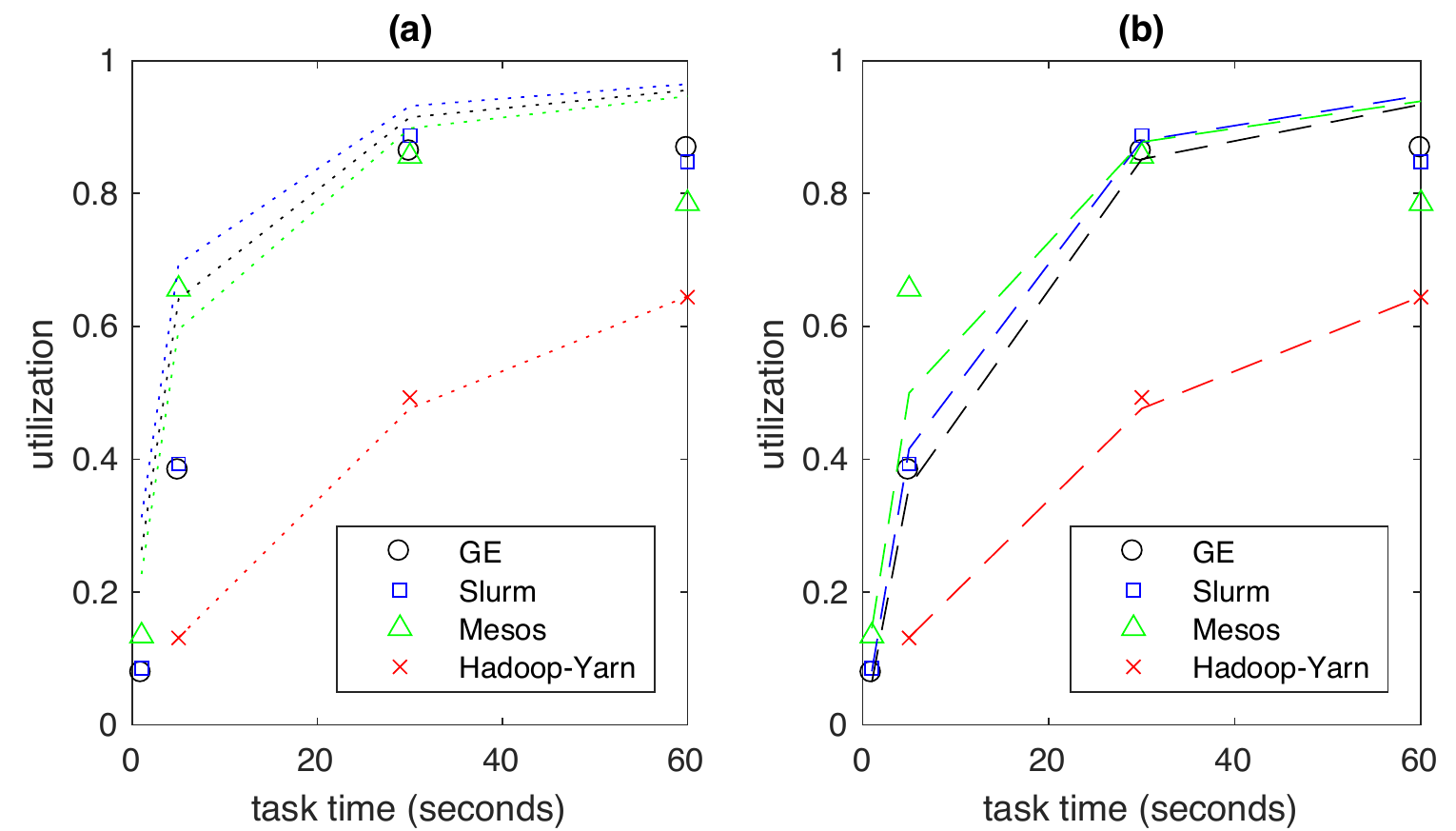}
    \caption{Scheduler utilization as a function of varying task times.  Points denote measured times.  (a) Dotted lines show approximate utilization model given by $U_c(t)^{-1} \approx 1 + t_s / t$.  (b) Dashed lines show  utilization model given by $U_c^{-1} = 1 + (t_s n^{\alpha_s}) / (t n)$.}
    \label{fig:Utilization1}
  \end{figure}

Figure~\ref{fig:Utilization1} shows the results of launch latency as measured by the utilization defined along with the approximate and more exact utilization models derived previously. In both cases, the models qualitatively agree with the data and indicate they can be used as reasonable approximations in theoretical utilization analyses. Since we have a fixed total amount of work, all additional time that it takes the cluster to execute the total set of jobs decreases the utilization of the cluster. Simply put, any time the scheduler consumes time in managing job execution (submission, queue management, resource identification, resource selection, resource allocation, job dispatch, and job termination) will decrease utilization. As we can see in Figure~\ref{fig:Utilization1}, all of the schedulers do well with 60-second tasks with the exception of Hadoop YARN; all of the schedulers launch jobs reasonably fast. Hadoop YARN has greater overhead for each job, including launching an application master process for each job~[\cite{white2015hadoop}]. Slurm, Grid Engine, and Mesos perform similarly with 1-, 5-, and 30-second tasks. However, Hadoop YARN is much less efficient among these schedulers; it was so inefficient that the 1-second task trials were prohibitive to run. Each of the schedulers loses efficiency as it must manage large queues of pending jobs, determine the best candidate job to launch next, and manage the launching and finishing of executing jobs. Besides the poor performance from Hadoop YARN, Slurm, Grid Engine, and Mesos show much lower utilization rates for 1- and 5-second jobs. This lackluster performance beg for a solution to realize better utilization. 

\subsection{Mulitlevel Scheduling}
The key to increasing the utilization for 1- and 5-second jobs is to decrease the job launch latency or not launch as many jobs overall while still getting all of the work done. With pleasantly parallel jobs, which most data analysis jobs are, the analysis code can be changed slightly to allow processing of multiple datasets or files with a single job execution. We can then use a tool like LLMapReduce [\cite{byun2016llmapreduce}] to efficiently run the jobs on the computing resource. This technique is referred to as multilevel scheduling. 

LLMapReduce is a tool that greatly simplifies the setup and execution of processing many data files with the same program. It employs the map-reduce parallel programming model, which consists of two user-written programs: a Mapper and a Reducer.  The input to the Mapper is a file, and the output is another file. The input to the Reducer is the set of Mapper output files. The output of Reducer is a single file. Launching consists of starting many Mapper programs, each with a different input file. When the Mapper programs all have completed, the Reduce program is run on the Mapper outputs. By default, LLMapReduce expects that the map application takes a single input and single output path (siso). However, this siso approach will incur overhead associated with repeated startups of the map application.  Some applications, such as MATLAB codes, can save significant overhead cost with the minor change of having the map application start only once and process many input/output file pairs. Using the multiple input and multiple output mode requires a (mildly) modified map application that will read the input file with the multiple lines of input/output filename pairs [\cite{byun2016llmapreduce}].

  \begin{figure}[!ht]
    \subfloat[Slurm]{
      \includegraphics[width=0.45\textwidth]{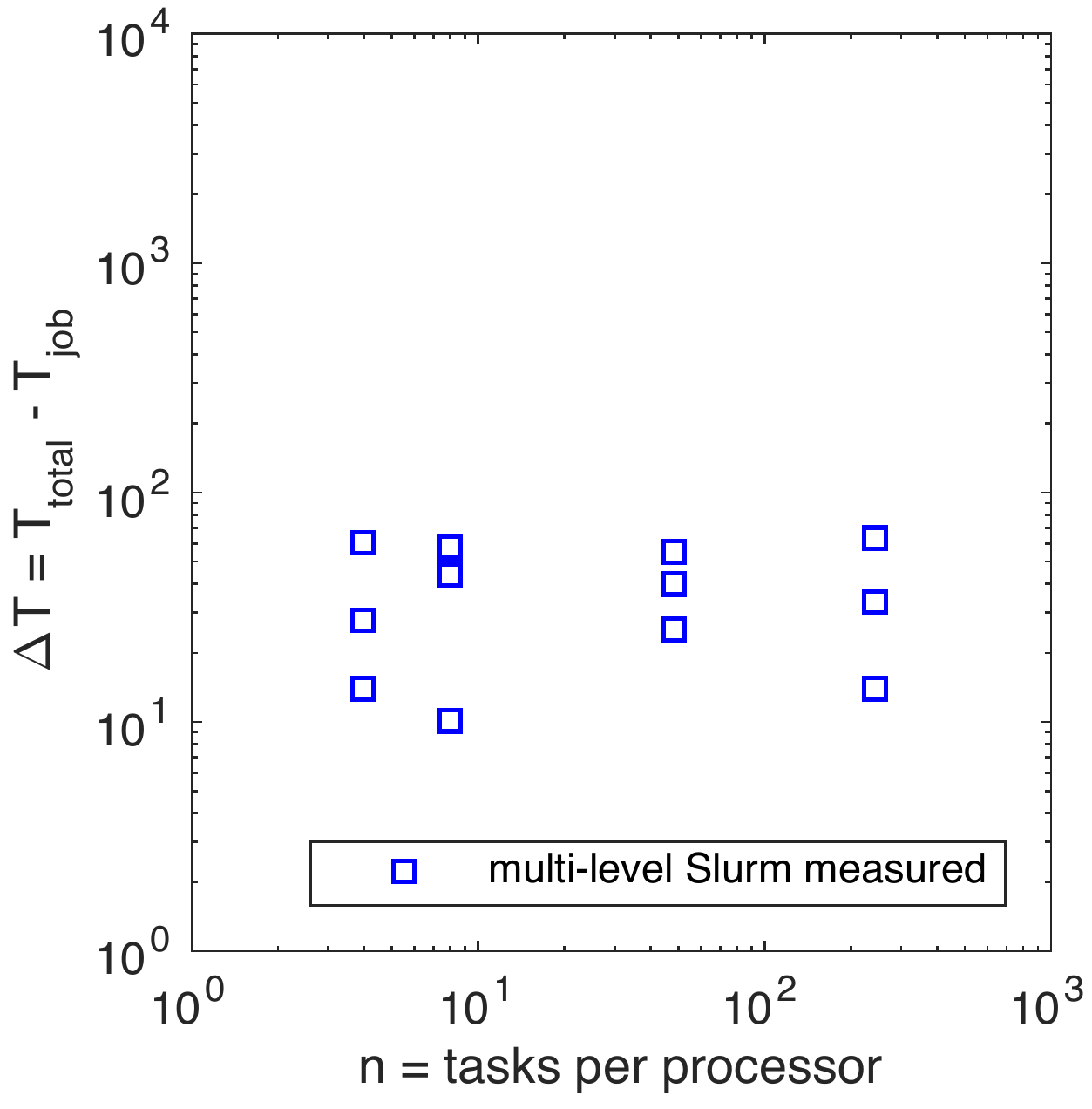}
      \label{subfig-1:LCam}
    }
    \hfill
    \subfloat[Grid Engine]{
      \includegraphics[width=0.45\textwidth]{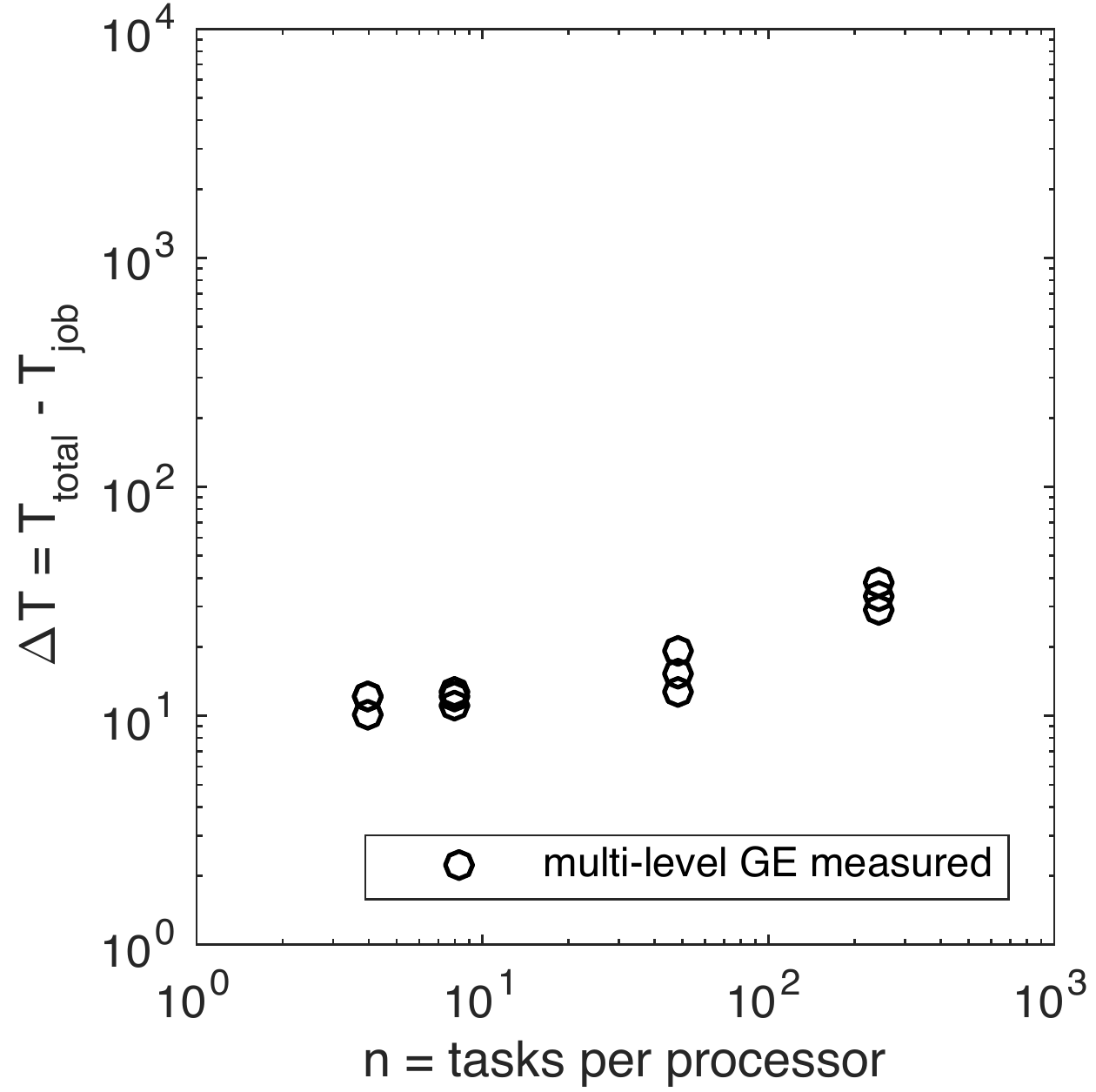}
       \label{subfig-2:LCbm}
    } \\
    \subfloat[Mesos]{
      \includegraphics[width=0.45\textwidth]{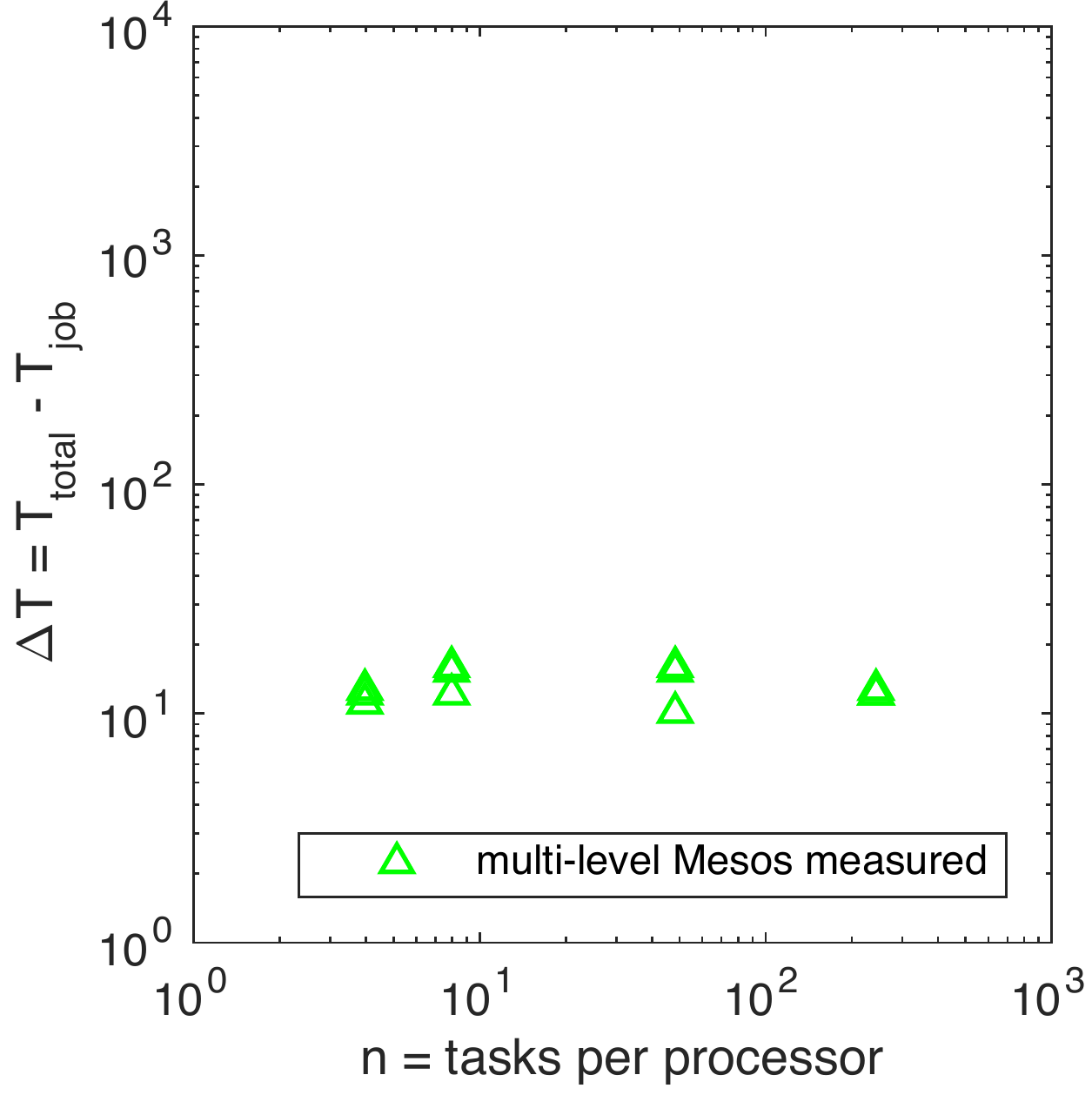}
      \label{subfig-3:LCcm}
    }
   \caption{Comparison of scheduler latency versus number of tasks per processor with multilevel scheduling.}
    \label{fig:LatencyComparisonMultilevel}
  \end{figure}

Again, we first analyzed the distribution of $\Delta T$ for each of the four schedulers under test as $n$, the number of tasks per processor, is increased. These results are shown in Figure~\ref{fig:LatencyComparisonMultilevel}. Again, note that these plots are on log-log axes.  Figure~\ref{subfig-1:LCam} shows significant scatter among trials at all values of $n$ with the Slurm scheduler with no apparent $\Delta T$ dependence upon $n$. Further, multilevel scheduling with Slurm reduces $\Delta T$ to below 100 seconds for all values of $n$, and there is a 30$\times$ reduction in $\Delta T$ at the largest values of $n$ when compared to not using multilevel scheduling. Grid Engine shows low scatter among trials at all values at $n$ in Figure~\ref{subfig-2:LCbm}, and there is a small increase in $\Delta T$ as $n$ increases. Here, multilevel scheduling reduces $\Delta T$ to below 40 seconds at all values of $n$; there is a 40$\times$ reduction for $\Delta T$ at the largest values of $n$ when compared to non-multilevel scheduling. Finally for multilevel scheduling with Mesos, Figure~\ref{subfig-3:LCcm} shows that there is low scatter among trials at all values of $n$, and there is no apparent $\Delta T$ dependence on $n$. Multilevel scheduling in Mesos reduces $\Delta T$ to below 20 seconds at all values of $n$, and there is a 100$\times$ reduction in $\Delta T$ at the largest values of $n$ when compared to non-multilevel scheduling. All of these results show that if pleasantly parallel computational jobs are amenable to using multilevel scheduling techniques, the effort to cast them as such is well worth it. 

\begin{figure}[htb]
    \centering
    \includegraphics[width=30pc]{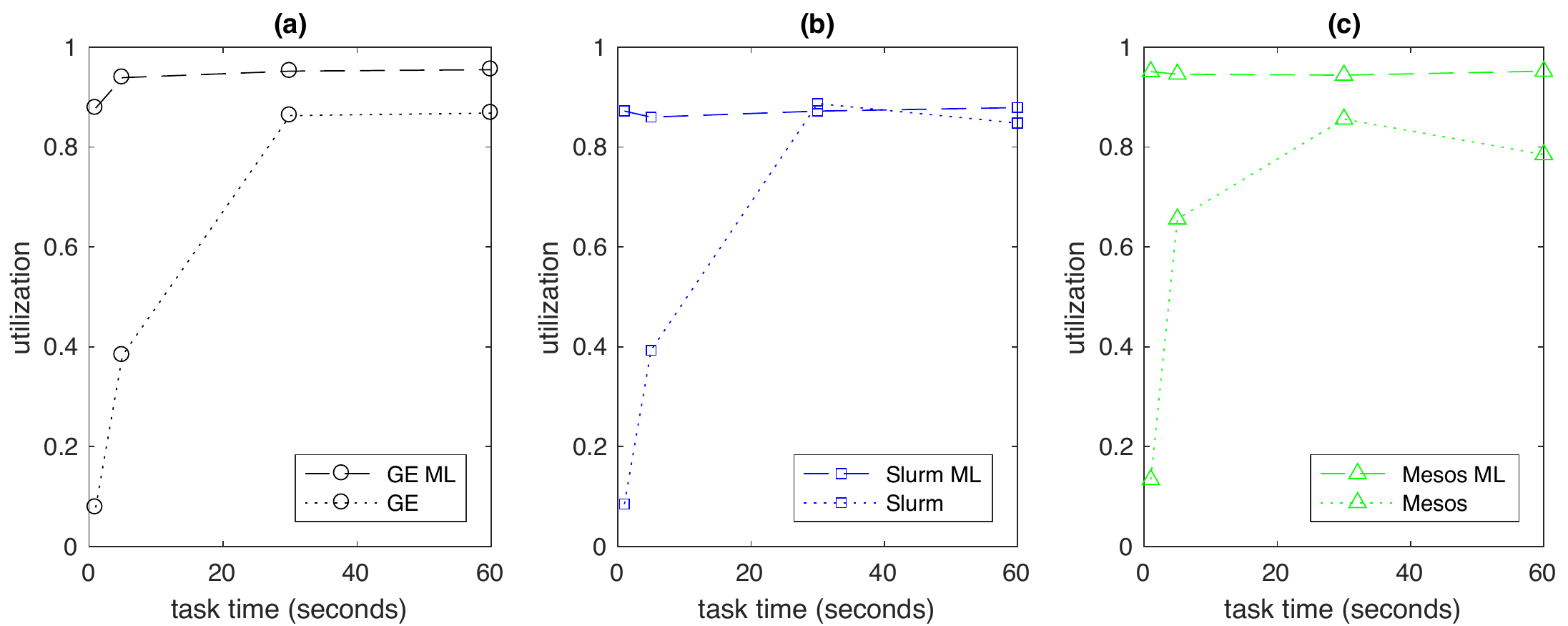}
    \caption{Utilization as function of tasks for regular and multilevel scheduling for (a) Grid Engine (b) Slurm, and (c) Mesos.  In all cases multilevel scheduling significantly improves utilization.}
    \label{fig:Utilization2}
  \end{figure}

The same four task sets from Table~\ref{tab:parameters1} in Subsection~\ref{subsec:latencyandutlization} were used with the LLMapReduce tool to test the utilization of multilevel scheduling on Slurm, Grid Engine, and Mesos. Figure~\ref{fig:Utilization2} shows the same results as Figure~\ref{fig:Utilization1} and includes the results of the use of multilevel scheduling. Slurm, Grid Engine, and Mesos all have high utilization with this technique. The figure shows that multilevel scheduling brings the utilization rates for all three schedulers around 90\%, which is on par with the 30- and 60-second jobs.

\section{Summary}
\label{sec:summary}
Schedulers are a primary component of a scalable computing infrastructure because they control all of the work on the system and directly impact overall system effectiveness. Schedulers have many features, and certain features can enhance the execution of certain workloads. In this paper, we have compared a number of scalable compute cluster schedulers and developed the concept of scheduler families based on their lineage and features. We then compared and contrasted a number of key features and their impact on high performance data analytic workloads. We found that, in general, there are many common features across all the schedulers that we examined. 

We then focused on scheduler latency, creating a scheduler latency model and validating the model a with a set of timing measurements on a subset of schedulers: Slurm, Grid Engine, Hadoop YARN, and Mesos. We found that Hadoop YARN performed worse than the other three schedulers and that the other three performed similarly. We then used the multilevel scheduling technique with the LLMapReduce tool to increase the overall utilization from Slurm, Grid Engine, and Mesos.



\section{Acknowledgments}
The authors wish to acknowledge the following individuals for their contributions: Alan Edelman, Justin Brukardt, Steve Pritchard, Chris Clarke, Sterling Foster, Paul Burkhardt, Victor Roytburd, Dave Martinez, Vijay Gadepally, Anna Klein, Lauren Milechin, Julie Mullen, Sid Samsi, and Chuck Yee.



\section*{References}
\bibliographystyle{elsarticle-harv} 
\bibliography{HPDASchedulersJDBC}


\end{document}